\documentclass[oldversion]{aa}

\usepackage{times}
\usepackage{graphicx}
\usepackage{xspace}
\usepackage{epsfig}
\usepackage{natbib}
\usepackage{lscape}

\newcommand{\mtr}{MTR\,13\xspace}

\def\gsim{\;\lower4pt\hbox{${\buildrel\displaystyle >\over\sim}$}\,}
\def\lsim{\;\lower4pt\hbox{${\buildrel\displaystyle <\over\sim}$}\,}

\begin{document}

\title{X-Shooter spectroscopy of young stellar objects\thanks{Based on observations collected at the Very Large Telescope of the European Southern Observatory under programs 084.C-0269, 085.C-0238, 086.C-0173, 087.C-0244, and 089.C-0143.}}

\subtitle{III. Photospheric and chromospheric properties of Class\,III objects}

\author{B. Stelzer \inst{1} \and A. Frasca \inst{2} \and J.M. Alcal\'a \inst{3}  
\and C.F. Manara \inst{4} \and K. Biazzo \inst{3,2} \and E. Covino \inst{3} \and 
E. Rigliaco \inst{5} \and L. Testi \inst{4} \and S. Covino \inst{6} \and V. D'Elia \inst{7}}

\offprints{B. Stelzer}

\institute{INAF - Osservatorio Astronomico di Palermo,
% INST 1
  Piazza del Parlamento 1,
  90134 Palermo, Italy \\ \email{B. Stelzer, stelzer@astropa.inaf.it} \and
% INST 2
  INAF - Osservatorio Astrofisico di Catania, 
  Via S.Sofia 78, 95123 Catania, Italy \and
% INST 3
  INAF - Osservatorio Astronomico di Capodimonte, 
  Via Moiariello 16, 
  80131 Napoli, Italy \and
% INST 4
  European Southern Observatory, Karl-Schwarzschild-Str.2, 85748 Garching, Germany \and
% INST 5
  Department of Planetary Science, Lunar and Planetary Lab, University of Arizona, 1629 E. University Blvd, 85719 Tucson, AZ, USA \and
% INST 6
  INAF - Osservatorio Astronomico di Brera, Via Bianchi 46, 23807 Merate, Italy \and 
% INST 7
  INAF - Osservaotorio Astronomico di Roma, Via di Frascati 33, 00040 Monte Porzio Catone, Italy % \and 
%% INST 8
%  INAF - Osservatorio Astrofisico di Arcetri,
%  Largo E. Fermi 5, 
%  50125 Firenze, Italy
}

\titlerunning{Photospheric and chromospheric properties of Class\,III objects}

\date{Received $<$28-05-2013$>$ / Accepted $<$30-07-2013$>$}

\abstract
{Traditionally, the chromospheres of late-type stars are studied through their strongest
emission lines, H$\alpha$ and \ion{Ca}{\sc ii}\,HK emission. Our knowledge on the whole
emission line spectrum is more elusive as a result of the limited spectral range and 
sensitivity of most available spectrographs.}
% context heading (optional)
{We intend to reduce this gap with a comprehensive spectroscopic study of the chromospheric
emission line spectrum of a sample of non-accreting pre-main sequence stars (Class\,III sources). 
}
% aims heading (mandatory)
{We analyzed X-Shooter/VLT spectra of $24$ Class\,III sources from three nearby 
star-forming regions ($\sigma$\,Orionis, Lupus\,III, and TW\,Hya). 
We determined the effective temperature, surface gravity, rotational 
velocity, and radial velocity by comparing the observed spectra with synthetic BT-Settl model 
spectra. We investigated
in detail the emission lines emerging from the stellar chromospheres and combined these
data with archival X-ray data to allow for a comparison between chromospheric and coronal 
emissions.
}
% methods heading (mandatory)
{ For some objects 
in the sample the atmospheric and kinematic parameters are presented here for the first time. 
The effective temperatures are consistent with those derived for the same stars from an 
empirical calibration with spectral types. Small differences in the surface gravity found between
the stars can be attributed to differences in the average age of the three star-forming regions. 
The strength of lithium absorption and radial velocities confirm the young age of all but
one object in the sample (Sz\,94). 
Both X-ray and H$\alpha$ luminosity as measured in terms of the bolometric luminosity 
are independent of the effective temperature for early-M stars but decline toward the 
end of the spectral M sequence. 
For the saturated early-M stars the average emission level is almost one dex
higher for X-rays than for H$\alpha$: $\log{(L_{\rm x}/L_{\rm bol})} = -2.85 \pm 0.36$ vs. 
$\log{(L_{\rm H\alpha}/L_{\rm bol})} = -3.72 \pm 0.21$. When all chromospheric emission
lines (including the Balmer series up to H11, \ion{Ca}\,{\sc II}\,HK, 
the \ion{Ca}\,{\sc II}\,infrared triplet, 
and several \ion{He}\,{\sc I} lines) are summed up the coronal flux still dominates that
of the chromosphere, typically by a factor $2-5$. Flux-flux relations between 
activity diagnostics that probe different atmospheric layers (from the lower chromosphere
to the corona) separate our sample of active pre-main sequence stars from the bulk
of field M dwarfs studied in the literature. Flux ratios between individual optical 
emission lines show a smooth dependence on the effective temperature. 
The Balmer decrements can roughly be reproduced by an NLTE radiative transfer model devised 
for another young star of similar age. Future, more complete chromospheric model grids can
be tested against this data set. 
}
% results heading (mandatory)
{}
% conclusions heading (optional)
%{
%}


\maketitle

\section{Introduction}\label{sect:intro}

The optical spectra of late-type stars are characterized by numerous emission lines. 
These lines are signatures of chromospheric activity that traces the reaction of the stellar
atmosphere to the magnetic processes related to the stellar dynamo. For FGK stars the
dynamo may be analogous to the one on the Sun, that is rooted in the interface between
radiative core and convective envelope \citep{Parker93.1}, but the nature 
of the dynamo in fully convective M stars is still elusive. In these stars 
the solar-like $\alpha\Omega$ dynamo may be replaced by turbulent magnetic fields 
\citep{Durney93.1} or by an $\alpha^2$ dynamo \citep{Chabrier06.1}. 
The same arguments hold for M stars on the pre-main sequence (PMS), which also have
fully convective interiors and are not expected to drive a solar-like interface dynamo.
Nevertheless, observations indicate that 
PMS stars manifest particularly strong signatures of chromospheric and coronal 
activity \citep[e.g.][]{Kuhi83.1,Walter86.1}. In fact, searching for coronal X-ray 
emission is a prime method for discovering PMS stars and, more recently, the UV emission
from the chromosphere and transition region has
been established as an analogous diagnostic for increasing the census of young stars 
\citep[e.g.][]{Findeisen10.0, Shkolnik11.0}. 
Studies of the physics of magnetic activity lag behind not least because all 
manifestations of magnetic activity (X-rays, UV radiation, optical emission lines) -- to
a lesser or larger extent -- present the problem that in PMS stars other 
processes such as accretion and outflows may be dominating the contribution from magnetic 
activity. PMS stars in the accreting stage (Class\,II sources or classical T\,Tauri stars)
have obtained more attention in the literature than non-accreting ones (Class\,III or 
weak-line T Tauri stars). 
A very small number of studies has been dedicated to chromospheric properties of 
Class\,III stars \citep[e.g.][]{Montes99.2}, and most of them either focused on 
a multi-wavelength study for an individual star \citep[e.g.][]{Welty95.1,Fernandez04.1} 
or treated a sample of stars, but for a single activity indicator, the H$\alpha$
line \citep{Scholz07.1}. 

H$\alpha$ is, indeed, the most widely studied emission feature in the optical spectra of
M stars. 
In contrast to FGK stars where the chromospheric contribution of H$\alpha$ is superposed
onto a photospheric absorption profile, H$\alpha$ has no absorption component according to 
synthetic model spectra in M dwarfs. This makes it easier to quantify the chromospheric 
activity. Moreover, limited wavelength coverage of the spectroscopic observations 
often makes H$\alpha$ the only available diagnostic for magnetic activity. 
For these reasons, H$\alpha$ is the traditional proxy for 
characterizing the level of magnetic activity on M stars. However, studies 
of activity in other wavelength bands, such as \ion{Ca}{\sc ii}\,H\&K emission, UV and
X-ray flux, have shown substantial emission in these activity signatures even in M dwarfs 
without H$\alpha$ emission \citep[e.g.][]{Walkowicz08.0, Houdebine11.0}. A comprehensive 
assessment of the chromospheric radiation budget, therefore, requires a study of multiple
activity diagnostics. 

An ideal instrument for this purpose is the X-Shooter spectrograph
of the {\it Very Large Telescope} at the European Southern Observatory. 
With its wavelength coverage from $3500-25000$\,\AA, it includes the whole
Balmer series, the calcium H\&K and infrared triplet (IRT) lines, several helium lines 
and the low-n lines of the Paschen and Brackett series. 
The various diagnostics for magnetic activity trace different layers of the outer 
atmospheres. The \ion{Ca}{\sc ii}\,H\&K and IRT line cores originate in the low/middle 
chromosphere, while the H$\alpha$ line carries information 
about upper layers, on average, although it forms in a wide atmospheric thickness
\citep[see e.g.][]{Vernazza81.0, Rutten07.0}. 
The \ion{He}{\sc i} lines are diagnostics of the upper chromosphere and lower
transition region, since the 
corresponding transitions require temperatures $\geq 10 000$\,K to be excited. 
The investigation of atmospheric structure can be expanded into the corona by means 
of X-ray observations \citep[e.g.][]{Vaiana81.1}. 

Here we present X-Shooter spectra of $24$ Class\,III PMS stars and X-ray data from 
{\em ROSAT} and {\em XMM-Newton} catalogs for the same stars. 
The Class\,III status of this sample was established 
on the basis of their spectral energy distribution and implies that they are
diskless and non-accreting such that the only process responsible for their line 
emission is magnetic activity. 
The main purpose of this work is to examine the relations between the emissions in the 
different lines to constrain the role of activity in the different
atmospheric layers. Moreover, we aim at constraining the dependence of the activity 
level on fundamental properties such as effective temperature ($T_{\rm eff}$), 
bolometric luminosity ($L_{\rm bol}$), rotation, and age, 
parameters that determine the nature of stellar dynamos. 
The radiation produced
through magnetic processes is known to be linked to the stellar bolometric luminosity and 
is often normalized to it by defining the `activity index', $L_{\rm proxy}/L_{\rm bol}$, 
where $L_{\rm proxy}$ is the luminosity of an emission line or a wavelength band 
(e.g. in the case of X-ray emission). 
Our sample spans the whole spectral M sequence and gives natural access to examining the 
role of $T_{\rm eff}$ from $2500$ to $4500$\,K. Bolometric luminosity, effective temperature, 
rotation rate ($v \sin{i}$), surface gravity, and lithium absorption are 
all determined using the same X-Shooter spectra as those in which we measured the activity 
diagnostics, which yields a self-consistent picture of photospheric and chromospheric 
properties of Class\,III sources to which we add the coronal perspective using archived 
X-ray data. 
Finally, we compare the results to chromospheric and coronal studies of main-sequence
M dwarfs from the literature to search for keys on the time evolution of magnetic activity. 

In Sect.~\ref{sect:data_analysis} we give more details on the sample and the data used. 
The determination of fundamental parameters, rotation, kinematics, and lithium absorption
and the analysis of the emission lines are described in Sect.~\ref{sect:analysis}.
The results on magnetic activity are presented in Sect.~\ref{sect:results}, and a 
summary and our conclusions are found in Sect.~\ref{sect:summary}.

\section{Sample and observations}\label{sect:data_analysis}

The sample considered in this paper comprises $24$ 
Class\,III objects, $13$ from the TW\,Hya association (henceforth TWA), 
$6$ from the Lupus\,III, and $5$ from the $\sigma$\,Orionis 
star-forming regions (SFRs). 
All stars were observed with X-Shooter within the 
INAF consortium's Guaranteed Time Observations (GTO)
\citep[see][for a presentation of the project]{Alcala11.0}.   
The target list is given in Table~\ref{tab:photosph_params}. The coordinates and
other designations for the objects are found in Table~1 of \cite{Manara13.0} 
(henceforth \mtr), where the
same sample and the data reduction are described in detail. We provide here only a
short summary. 

The classification as Class\,III sources was derived using published
{\em Spitzer} photometry and spectroscopic features. 
In particular, we made use of the slopes of the spectral energy distribution (SEDs) 
determined by \cite{Merin08.1} for Lupus and by \cite{Hernandez07.1} for $\sigma$\,Ori
members. Because the mid- and far-IR SEDs are not available or incomplete for most stars
in the TWA, we have used for these stars the H$\alpha$ equivalent width criterion defined
by \cite{White03.1} and the absence of forbidden emission lines in our X-Shooter spectra. 
The H$\alpha$ and the forbidden lines were also used to
confirm the Class\,III status of the objects in Lupus and $\sigma$\,Ori.

\subsection{X-Shooter spectra}\label{subsect:data_xshooter}

The X-Shooter observations were carried out between May 2010 and April 2012.
Individual exposure times and slit widths were chosen depending on the
brightness of the star 
%\citep[see Table~2 of ][]{Manara13.0}. 
(see Table~2 of \mtr). Briefly, the  
exposure times ranged from $100$\,sec to $3600$\,sec, and the slit widths yielded 
spectral resolutions ranging from $R \sim 3300...9100$ in the UVB,
$R \sim 5400...17400$ in the VIS, and $R \sim 3500...11300$ in the NIR. 
The data reduction was performed with the X-Shooter pipeline \citep{Modigliani10.0}.
Finally, the flux-calibrated spectrum was scaled by an individual factor for 
each star to adapt it to the published broad-band photometry. In this way, eventual slit 
losses and non-photometric conditions were compensated for. 

%\cite{Manara13.0}
\mtr have defined a spectral sequence for this sample of Class\,III
objects by ordering the stars according to the depth of molecular features. 
Some widely used spectral indices were shown to be consistent with this scale.
In Table~\ref{tab:photosph_params} we report some parameters derived and/or compiled by 
%\cite{Manara13.0} 
\mtr that are useful for our analysis: 
the spectral types (SpT), the bolometric luminosities ($L_{\rm bol}$), 
and the distances ($d$). 
We also give the values for the photometric rotation periods from the literature 
($P_{\rm rot}$). 
\begin{table*}\begin{center}
\caption{Fundamental and atmospheric parameters, rotation and kinematics of Class\,III sample.}
\label{tab:photosph_params}
\begin{tabular}{llrrr|ccrcc|c}\hline
\multicolumn{1}{c}{Name} & \multicolumn{1}{c}{SpT$^{(\dagger)}$} & \multicolumn{1}{c}{${\log{\frac{L_{\rm bol}}{L_\odot}}}^{(\dagger)}$} & \multicolumn{1}{c}{$d^{(\dagger)}$} & \multicolumn{1}{c}{${P_{\rm rot}}^{(\ddagger)}$} & \multicolumn{1}{c}{$T_{\rm eff}$} & \multicolumn{1}{c}{$\log{g}$} & \multicolumn{1}{c}{$R_*$} & \multicolumn{1}{c}{$v \sin{i}$} & \multicolumn{1}{c}{$RV$} & \multicolumn{1}{c}{$W_{\rm Li}$} \\
 & & & [pc] & \multicolumn{1}{c}{[d]} & [K] & [${\rm cm/s^2}$] & [$R_\odot$] & [km/s] & [km/s] & [m\AA] \\ \hline
             TWA\,9A &   K5  & $ -0.61$ & $   68$ & $5.10$ & $ 4333 \pm  47$ & $   4.9  \pm   0.8$ & $  0.87$ & $  14.2 \pm    0.4$ & $ 11.7 \pm   0.4$ & $  498 \pm    53$ \\
             SO\,879 &   K7  & $ -0.29$ & $  360$ & $ ...$ & $ 4021 \pm  85$ & $   3.9  \pm   0.5$ & $  1.47$ & $  10.0 \pm    4.0$ & $ 35.0 \pm   1.3$ & $  599 \pm    66$ \\
              TWA\,6 &   K7  & $ -0.96$ & $   51$ & $0.54$ & $ 4020 \pm 173$ & $   4.8  \pm   0.5$ & $  0.68$ & $  76.8 \pm    1.3$ & $ 17.1 \pm   1.5$ & $  485 \pm    54$ \\
             TWA\,25 &   M0  & $ -0.61$ & $   54$ & $ ...$ & $ 3749 \pm  60$ & $   4.7  \pm   0.5$ & $  1.17$ & $  14.4 \pm    0.5$ & $  8.8 \pm   1.2$ & $  549 \pm    58$ \\
             TWA\,14 & M0.5  & $ -0.83$ & $   96$ & $0.63$ & $ 3756 \pm  99$ & $   4.7  \pm   0.6$ & $  0.90$ & $  46.0 \pm    0.7$ & $ 14.9 \pm   1.3$ & $  577 \pm    59$ \\
            TWA\,13B &   M1  & $ -0.70$ & $   59$ & $5.35$ & $ 3661 \pm 103$ & $   4.6  \pm   0.6$ & $  1.10$ & $  10.4 \pm    1.1$ & $ 11.2 \pm   0.7$ & $  526 \pm    57$ \\
            TWA\,13A &   M1  & $ -0.61$ & $   59$ & $5.56$ & $ 3616 \pm  87$ & $   4.8  \pm   0.5$ & $  1.25$ & $  10.0 \pm    0.7$ & $ 10.0 \pm   1.0$ & $  595 \pm    65$ \\
             TWA\,2A &   M2  & $ -0.48$ & $   47$ & $ ...$ & $ 3533 \pm 133$ & $   4.4  \pm   0.4$ & $  1.53$ & $  18.0 \pm    1.9$ & $ 10.0 \pm   1.1$ & $  536 \pm    63$ \\
             Sz\,122$^{(*)}$ &   M2  & $ -0.60$ & $  200$ & $ ...$ & $ 3494 \pm 126$ & $   4.6  \pm   0.5$ & $  1.36$ & $ 150.2 \pm    2.7$ & $ 18.1 \pm   3.0$ & $  268 \pm    54$ \\
             TWA\,9B &   M3  & $ -1.17$ & $   68$ & $3.98$ & $ 3357 \pm  67$ & $   4.5  \pm   0.6$ & $  0.76$ & $  11.6 \pm    2.2$ & $ 11.0 \pm   0.5$ & $  476 \pm    57$ \\
            TWA\,15B &   M3  & $ -0.96$ & $  111$ & $0.72$ & $ 3463 \pm 129$ & $   4.6  \pm   0.6$ & $  0.91$ & $  21.6 \pm    0.9$ & $ 11.8 \pm   1.2$ & $  477 \pm    65$ \\
              TWA\,7 &   M3  & $ -1.14$ & $   28$ & $5.05$ & $ 3408 \pm  85$ & $   4.4  \pm   0.4$ & $  0.77$ & $   8.4 \pm    1.3$ & $  8.4 \pm   0.8$ & $  551 \pm    64$ \\
            TWA\,15A & M3.5  & $ -0.95$ & $  111$ & $0.65$ & $ 3347 \pm 125$ & $   4.5  \pm   0.3$ & $  0.99$ & $  33.6 \pm    0.9$ & $ 10.1 \pm   1.7$ & $  542 \pm    62$ \\
             Sz\,121$^{(*)}$ &   M4  & $ -0.34$ & $  200$ & $ ...$ & $ 3286 \pm 134$ & $   4.1  \pm   0.3$ & $  2.07$ & $  84.8 \pm    3.3$ & $ 13.6 \pm   2.5$ & $  575 \pm    68$ \\
              Sz\,94 &   M4  & $ -0.76$ & $  200$ & $ ...$ & $ 3219 \pm 101$ & $   4.3  \pm   0.3$ & $  1.33$ & $  29.8 \pm    2.7$ & $  6.4 \pm   2.6$ & $...$ \\
             SO\,797 & M4.5  & $ -1.26$ & $  360$ & $ ...$ & $ 3274 \pm 136$ & $   3.9  \pm   0.4$ & $  0.72$ & $  27.4 \pm    0.9$ & $ 32.0 \pm   2.0$ & $  548 \pm    67$ \\
             SO\,641 &   M5  & $ -1.53$ & $  360$ & $1.75$ & $ 3176 \pm 140$ & $   3.8  \pm   0.4$ & $  0.56$ & $  12.5 \pm    4.9$ & $ 29.5 \pm   3.8$ & $  554 \pm    69$ \\
          Par-Lup3-2 &   M5  & $ -0.75$ & $  200$ & $ ...$ & $ 3038 \pm 101$ & $   3.7  \pm   0.4$ & $  1.51$ & $  26.4 \pm    4.7$ & $  4.4 \pm   3.7$ & $  575 \pm    72$ \\
             SO\,925 & M5.5  & $ -1.59$ & $  360$ & $1.79$ & $ 3113 \pm 178$ & $   3.8  \pm   0.4$ & $  0.55$ & $  20.8 \pm    4.2$ & $ 31.7 \pm   4.4$ & $  530 \pm    78$ \\
             SO\,999 & M5.5  & $ -1.28$ & $  360$ & $0.95$ & $ 3047 \pm 131$ & $   3.8  \pm   0.4$ & $  0.82$ & $  16.3 \pm    4.0$ & $ 31.2 \pm   2.0$ & $  593 \pm    84$ \\
             Sz\,107 & M5.5  & $ -0.79$ & $  200$ & $ ...$ & $ 3045 \pm 111$ & $   3.7  \pm   0.3$ & $  1.44$ & $  72.5 \pm    3.3$ & $  3.6 \pm   4.8$ & $  532 \pm    89$ \\
          Par-Lup3-1 & M6.5  & $ -1.18$ & $  200$ & $ ...$ & $ 2860 \pm 106$ & $   3.6  \pm   0.4$ & $  1.04$ & $  19.0 \pm    2.0$ & $  4.9 \pm   4.4$ & $  529 \pm   177$ \\
             TWA\,26 &   M9  & $ -2.70$ & $   42$ & $ ...$ & $ 2476 \pm 109$ & $   3.6  \pm   0.5$ & $  0.24$ & $  34.9 \pm   14.3$ & $  5.2 \pm   9.8$ & $  578 \pm    99$ \\
             TWA\,29 & M9.5  & $ -2.81$ & $   79$ & $ ...$ & $ 2390 \pm 109$ & $   3.6  \pm   0.3$ & $  0.23$ & $  35.8 \pm   17.2$ & $  7.7 \pm   3.9$ & $  590 \pm   133$ \\
\hline
\multicolumn{11}{l}{$^{(*)}$ Fundamental and kinematic parameters of Sz\,121 and Sz\,122 are uncertain due to possible binarity.} \\
\multicolumn{11}{l}{$^{(\dagger)}$ Spectral type, bolometric luminosity and distance adopted from \mtr.} \\
\multicolumn{11}{l}{$^{(\ddagger)}$ Rotation periods from \protect\cite{Lawson05.1} for TWA members and from \protect\cite{Cody10.0} for $\sigma$\,Ori members.} \\
\end{tabular}
\end{center}\end{table*}

An essential part throughout this study is the analysis of spectral lines, and we
anticipate here the treatment of the uncertainties of the line fluxes.
We considered them to be composed of two components, the statistical measurement error 
and a systematic error associated with the flux calibration.
%The former ones have been determined by \mtr ....{\it Carlo? How did you do it?}. 
As mentioned above, \mtr have normalized the spectra to published photometry such that 
the error in the flux eventually traces the uncertainties of the photometric measurements 
and especially the variations due to magnetic activity which we assumed to be 
$10$\,\%. (We note that the multi-band photometry of our objects is not contemporaneous.)  
The two uncertainties (statistical and systematic) are summed quadratically.

\subsection{Ancillary X-ray data}\label{subsect:data_anc}

We have collected X-ray data from the literature and from public data archives.
The flux measured at Earth is given for all stars in 
%Table~\ref{tab:xray_params} 
Table~\ref{tab:chromo_fluxes} 
and was obtained in the following way. 
\begin{table*}\begin{center}
\caption{Coronal and chromospheric fluxes. X-ray fluxes ($f_{\rm X}$) and upper/lower bounds with references from the literature (see Sect.~\ref{subsect:data_anc}). Optical chromospheric flux ($f_{\rm opt}$) from the sum of the emission lines in the X-Shooter spectra. FUV and NUV fluxes from the literature (see Sect.~\ref{subsect:results_fluxflux}).}
\label{tab:chromo_fluxes}
\begin{tabular}{lrcclrcccc}\hline
\multicolumn{1}{c}{Name} & \multicolumn{3}{c}{$\log{f_{\rm X}}$} & Ref. & \multicolumn{2}{c}{$\log{f_{\rm opt}}$} & \multicolumn{1}{c}{$\log{f_{\rm FUV,HST}}$} & \multicolumn{1}{c}{$\log{f_{\rm FUV,GALEX}}$} & \multicolumn{1}{c}{$\log{f_{\rm NUV,GALEX}}$} \\
\multicolumn{1}{c}{} & \multicolumn{3}{c}{${\rm [erg/cm^2/s]}$}  &      & \multicolumn{2}{c}{${\rm [erg/cm^2/s]}$} & \multicolumn{1}{c}{${\rm [erg/cm^2/s]}$} & \multicolumn{1}{c}{${\rm [erg/cm^2/s]}$} & \multicolumn{1}{c}{${\rm [erg/cm^2/s]}$} \\ \hline
             TWA\,9A & $ $ & $  -11.51$ & $ [-11.61,-11.44]$ &       & $ $ & $  -12.17$ & $  -12.85$ & $     ...$ & $     ...$ \\
             SO\,879 & $ $ & $  -13.08$ & $ [-13.13,-13.03]$ &   (1) & $ $ & $  -13.13$ & $     ...$ & $     ...$ & $     ...$ \\
              TWA\,6 & $ $ & $  -11.58$ & $ [-11.65,-11.52]$ &       & $ $ & $  -12.02$ & $  -12.57$ & $     ...$ & $     ...$ \\
             TWA\,25 & $ $ & $  -11.47$ & $ [-11.57,-11.40]$ &       & $ $ & $  -11.73$ & $     ...$ & $     ...$ & $     ...$ \\
             TWA\,14 & $ $ & $  -11.97$ & $ [-12.07,-11.88]$ &       & $ $ & $  -12.40$ & $     ...$ & $  -12.32$ & $  -12.57$ \\
            TWA\,13B & $ $ & $  -11.42$ & $ [-11.48,-11.37]$ &       & $ $ & $  -12.03$ & $  -12.91$ & $  -12.68$ & $  -12.64$ \\
            TWA\,13A & $ $ & $  -11.42$ & $ [-11.48,-11.37]$ &       & $ $ & $  -11.63$ & $  -12.89$ & $  -12.68$ & $  -12.64$ \\
             TWA\,2A & $ $ & $  -11.52$ & $ [-11.58,-11.46]$ &       & $ $ & $  -11.82$ & $     ...$ & $  -12.82$ & $  -12.71$ \\
             Sz\,122 & $ $ & $  -12.73$ & $ [-12.96,-12.58]$ &       & $ $ & $  -12.62$ & $     ...$ & $     ...$ & $     ...$ \\
             TWA\,9B & $ $ & $  -11.51$ & $ [-11.61,-11.44]$ &       & $ $ & $  -12.77$ & $  -13.84$ & $     ...$ & $     ...$ \\
            TWA\,15B & $ $ & $  -11.99$ & $ [-12.08,-11.91]$ &       & $ $ & $  -12.74$ & $     ...$ & $     ...$ & $     ...$ \\
              TWA\,7 & $ $ & $  -11.54$ & $ [-11.60,-11.48]$ &       & $ $ & $  -11.86$ & $  -12.58$ & $     ...$ & $     ...$ \\
            TWA\,15A & $ $ & $  -11.99$ & $ [-12.08,-11.91]$ &       & $ $ & $  -12.58$ & $     ...$ & $     ...$ & $     ...$ \\
             Sz\,121 & $<$ & $  -12.75$ & $                $ &       & $<$ & $  -12.94$ & $     ...$ & $     ...$ & $     ...$ \\
              Sz\,94 & $ $ & $  -13.39$ & $ [-13.44,-13.35]$ &   (2) & $ $ & $  -13.22$ & $     ...$ & $     ...$ & $     ...$ \\
             SO\,797 & $ $ & $  -14.34$ & $ [-14.47,-14.25]$ &   (1) & $ $ & $  -14.47$ & $     ...$ & $     ...$ & $     ...$ \\
             SO\,641 & $<$ & $  -12.75$ & $                $ &       & $ $ & $  -14.75$ & $     ...$ & $     ...$ & $     ...$ \\
          Par-Lup3-2 & $ $ & $  -13.68$ & $ [-13.73,-13.63]$ &   (2) & $<$ & $  -13.65$ & $     ...$ & $     ...$ & $     ...$ \\
             SO\,925 & $<$ & $  -12.75$ & $                $ &       & $<$ & $  -14.81$ & $     ...$ & $     ...$ & $     ...$ \\
             SO\,999 & $<$ & $  -12.75$ & $                $ &       & $ $ & $  -14.44$ & $     ...$ & $     ...$ & $     ...$ \\
             Sz\,107 & $ $ & $  -13.71$ & $ [-13.76,-13.66]$ &   (2) & $ $ & $  -13.41$ & $     ...$ & $     ...$ & $     ...$ \\
          Par-Lup3-1 & $ $ & $  -13.41$ & $ [-13.46,-13.37]$ &   (2) & $<$ & $  -14.43$ & $     ...$ & $     ...$ & $     ...$ \\
             TWA\,26 & $ $ & $  -15.25$ & $ [-15.73,-15.03]$ &   (3) & $<$ & $  -14.96$ & $     ...$ & $     ...$ & $     ...$ \\
             TWA\,29 & $<$ & $  -14.63$ & $                $ &   (4) & $<$ & $  -15.29$ & $     ...$ & $     ...$ & $     ...$ \\
\hline
\multicolumn{10}{l}{References: (1) - \protect\cite{Franciosini06.1}, (2) - \protect\cite{Gondoin06.1}, (3) - \protect\cite{Castro11.0}, (4) - \protect\cite{Stelzer07.2}.} \\
\end{tabular}
\end{center}\end{table*}

In a first step, we cross-correlated the target list with the Bright Source
Catalog \citep[BSC;][]{Voges99.1} 
and with the Faint Source Catalog \citep[FSC;][]{Voges00.1a} 
of the {\em ROSAT} All-Sky Survey (RASS). We used a match radius
of $40^{\prime\prime}$ \citep[e.g.][]{Neuhaeuser95.1}. 
We found that $11$ stars are identified with a BSC X-ray source and
one is detected in the FSC. 
To obtain X-ray fluxes for the other half of our sample we searched the literature. 
For the stars in $\sigma$\,Ori and Lupus we made use of the {\em XMM-Newton}
measurements published by \cite{Franciosini06.1} and \cite{Gondoin06.1}, respectively.
Four of the six Class\,III objects in Lupus and two of the five in $\sigma$\,Ori
are listed as X-ray sources in these works.  
The TWA covers a large area on the sky and no dedicated {\em XMM-Newton} or {\em Chandra}
observations have been performed to study its stellar population. 
We note that $11$ of the $12$ RASS identifications are objects from the TWA. 
The two brown dwarfs (BDs), \object{TWA\,26} and 
\object{TWA\,29}, are not detected in the RASS but 
their X-ray emission has been studied by \cite{Castro11.0} and \cite{Stelzer07.2}
on the basis of a dedicated {\em Chandra} and a serendipitous {\em XMM-Newton}
observation, respectively. 
\object{TWA\,29} was not detected and an upper limit
was given by \cite{Stelzer07.2}. For the remaining four targets without X-ray detection
we assumed an upper limit to the RASS count rate of $0.02$\,cts/s, corresponding to
the sensitivity limit for a typical RASS exposure time of $\approx 400-500$\,sec
\citep[see][]{Stelzer13.0}. 
The matches and table handling was carried out in the Virtual Observatory environment
{\sc TOPCAT} \citep{Taylor05.0} and in {\sc IDL}\footnote{The Interactive Data Language (IDL)
is a registered trademark of ITT Visual Information Solutions.}.

We transformed the RASS count rates and upper limits 
into $0.12-2.5$\,keV flux assuming a thermal model with 
plasma temperature of $kT = 1$\,keV 
and an absorbing column density of $N_{\rm H} = 10^{20}\,{\rm cm^{-2}}$. 
Then, X-ray luminosities were calculated using the individual distances 
from Table~\ref{tab:photosph_params}. For the stars with X-ray data in the literature
the published X-ray luminosities can be used. However, to make them comparable with those 
from the RASS, two corrections had to be applied. The first one is a transformation
from the {\em XMM-Newton} and {\em Chandra} energy bands for which the luminosities
had been calculated to the energy band of {\em ROSAT}.
The correction factors for the literature sources are between $0.99$ and $1.38$. 
The published X-ray luminosities were multiplied by these factors. Moreover, they  
were corrected for the differences in the distances adopted in the literature
and by us. 

We added a $10$\,\% systematic error to the statistical errors of all X-ray measurements.
The lower and upper values for the X-ray flux corresponding to the uncertainties
are given in brackets in Table~\ref{tab:chromo_fluxes}.

\section{Analysis}\label{sect:analysis}

\subsection{Stellar parameters and rotation determined with ROTFIT}\label{subsect:analysis_rotfit}

We used the ROTFIT code \citep{Frasca06.1} for evaluating the atmospheric parameters, 
effective temperature ($T_{\rm eff}$) and surface gravity ($\log g$),    
and for determining the projected rotation velocity 
($v\sin i$). For this purpose, we adopted as templates a grid of synthetic BT-Settl 
spectra \citep{Allard10.0} with solar metallicity, effective temperature in the 
range 2300--4400\,K (in steps of 100\,K) and $\log g$ from 5.5 to 3.0 (in steps of 0.5).
The synthetic spectra were degraded in resolution to match that of the target 
spectra, 
which depends on the adopted slit width, and they were resampled on their spectral points.
For each target spectrum, ROTFIT sorts the synthetic templates according to the $\chi^2$ 
of the residuals ``observed - template", where the synthetic templates are progressively 
broadened by convolution with a rotational profile of increasing $v\sin i$ until a minimum 
$\chi^2$ is reached. The stellar parameters
adopted by us are averages of the values of the five best templates (for each 
of the analyzed spectral regions described in detail below) 
weighted according to the $\chi^2$. The standard error of 
the weighted mean was adopted as the measurement error.
By using the five best-fitting models instead of the one best fit 
we achieve a higher accuracy because of the coarse sampling of the stellar parameters 
in the grid of templates.
The weighted average with a larger number of model spectra (e.g. ten) does not 
change significantly since the $\chi^2$ increases quite rapidly after the first 
five models, lowering their weight. 

The analysis is based on the assumption of negligible extinction in our targets. 
In fact, the three star-forming regions observed during our X-Shooter survey were chosen,
amongst others, because their low extinction yields a high signal in the UVB range where
very low-mass YSOs are intrinsically faint. Individual values for the optical extinction, 
$A_{\rm V}$, of Lupus members can be found in \cite{Hughes94.1} and \cite{Comeron03.1}.
For all but one object of our Lupus sample $A_{\rm V} = 0$\,mag. 
For the low reddening of the $\sigma$\,Ori cluster see for instance the references in 
\cite{Walter08.0}. 
Finally, the TWA is a nearby young association with negligible extinction because
it is located far from molecular clouds. 

The derived values for $T_{\rm eff}$, $\log{g}$ and $v\sin{i}$ are given in 
Table~\ref{tab:photosph_params}. 
We also list the stellar radii ($R_*$) computed from the effective temperatures and the 
bolometric luminosities with the Stefan-Boltzmann law.

\subsubsection{Spectral regions for ROTFIT analysis}\label{subsubsect:analysis_rotfit_regions}

\begin{figure}
\begin{center}
\parbox{9cm}{
\includegraphics[width=9cm]{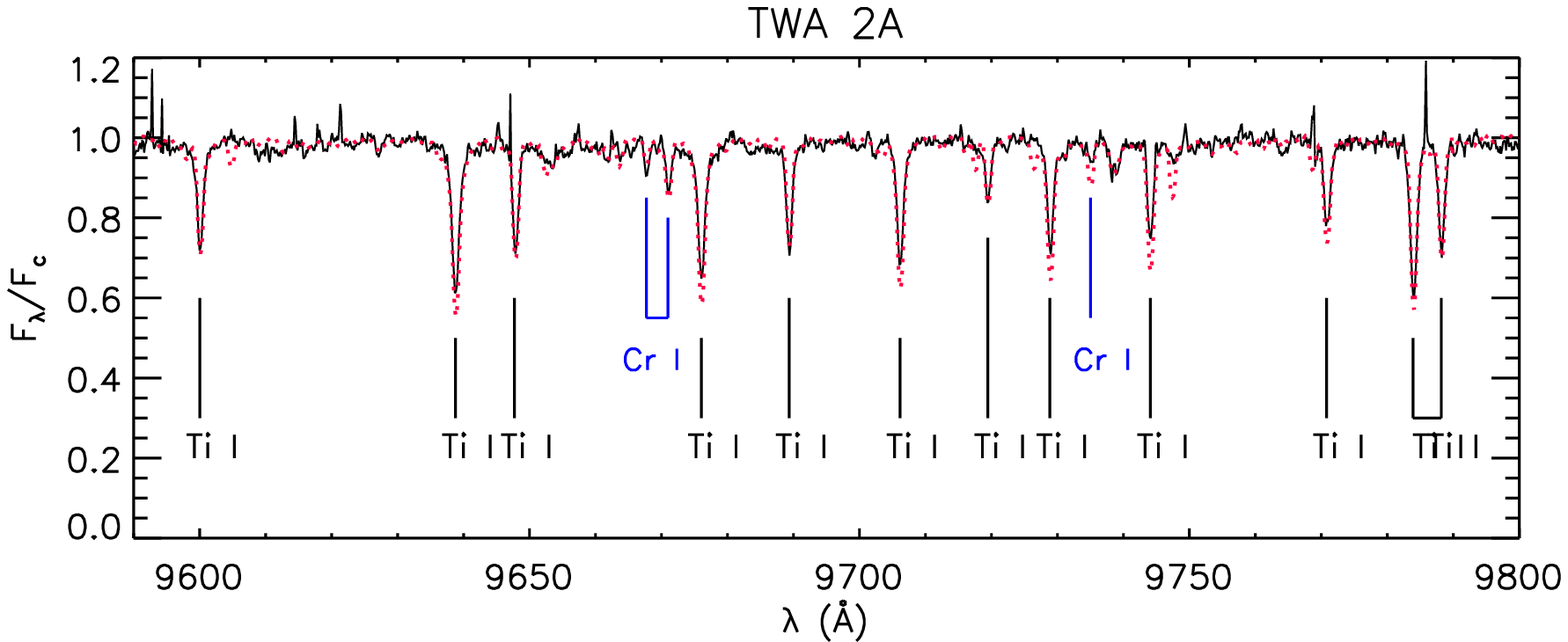}
\includegraphics[width=9cm]{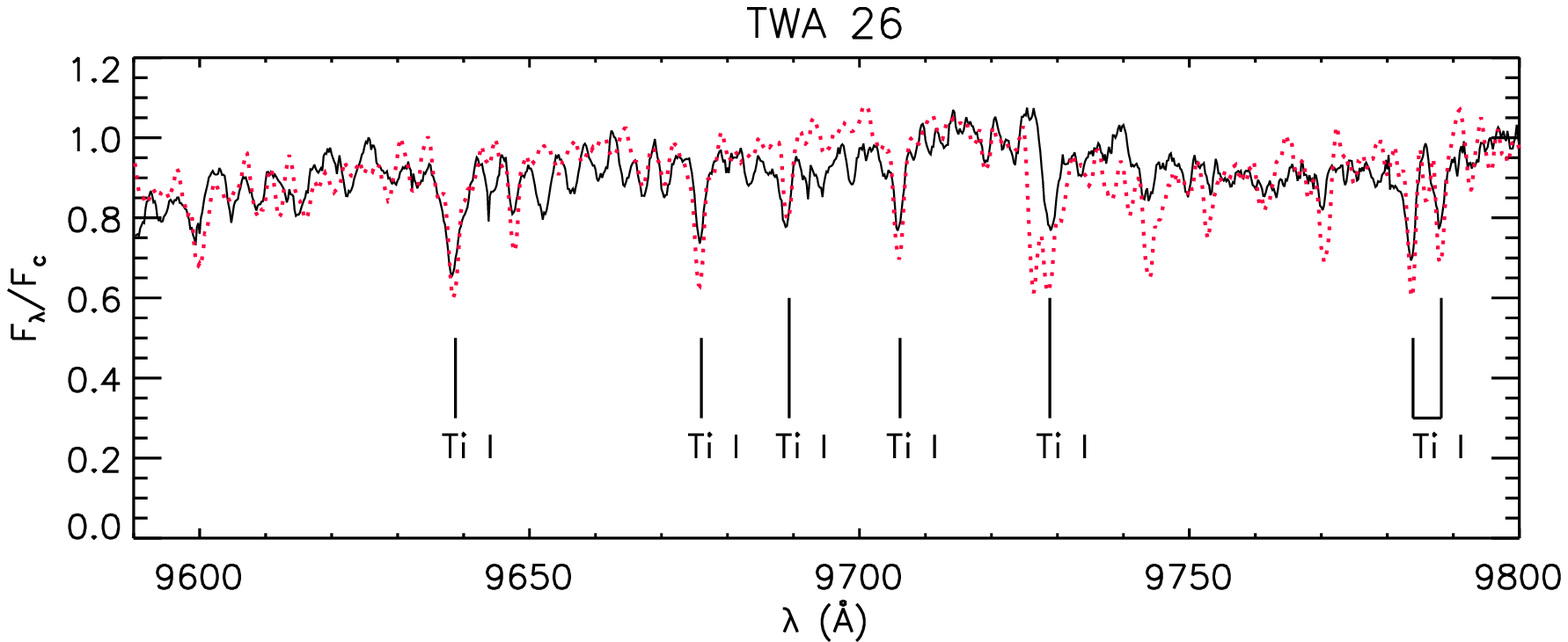}
}
\caption{Portion of continuum-normalized X-Shooter VIS spectra of \object{TWA\,2A}
%CD-29\,8887A 
({\it top panel}) and \object{TWA\,26} ({\it lower panel}) around $9700$\,\AA~(solid 
black lines). 
The best-fitting synthetic spectra, properly Doppler-shifted and rotationally broadened, 
are overplotted (dotted red lines). The main absorption lines are also marked.}
\label{fig:lambda9700}
\end{center}
\end{figure}

The best spectral regions for performing an accurate determination of $v\sin i$ with 
ROTFIT are those free of very broad lines and strong molecular bands, but including 
several absorption lines strong enough to still be prominent features against a 
well-defined continuum in fast rotators as well. 
We found a very well-suited region between 9600 and 9800\,\AA\ where several sharp absorption 
lines (mainly of \ion{Ti}{i} and \ion{Cr}{I}) are present (c.f. Fig\,\ref{fig:lambda9700}) 
for stars hotter than 2900\,K. Moreover, the intensity of these lines scales with 
$T_{\rm eff}$ as apparent from the BT-Settl models, 
thus they can be also used as temperature diagnostics. We derived the 
$v\sin i$ as the weighted mean described above using only this spectral region and
kept it fixed inside ROTFIT to determinate the atmospheric parameters  
based on other spectral diagnostics. 

\begin{figure}
\begin{center}
\parbox{9cm}{
\includegraphics[width=9cm]{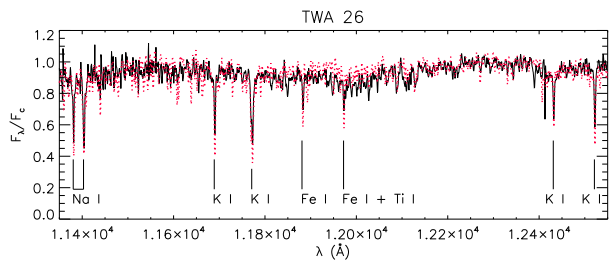}
}
\caption{Portion of the X-Shooter NIR spectrum of \object{TWA\,26} around 
$12 000$\,\AA~(solid black line) and the best fitting synthetic spectrum 
(dotted red line).} 
\label{fig:lambda12000}
\end{center}
\end{figure}

The spectral region around 9700\,\AA\, of the two coolest objects in our sample, 
\object{TWA\,26} and \object{TWA\,29}, is much more 
affected by molecular bands and noise than that of the hotter stars, and the 
\ion{Ti}{i} absorption lines become quite faint (see Fig.~\ref{fig:lambda9700}). 
Thus we used another 
%we were forced to use another 
spectral region as temperature diagnostics, 
the wavelength range $11300-12600$\,\AA\, in the NIR spectra,   
where several absorption lines are visible (c.f. Fig.\,\ref{fig:lambda12000}). 
Unfortunately, the resolution and sampling of the NIR spectra is lower than 
that of the VIS spectra, which makes the $v\sin i$ determinations less accurate. 

\begin{figure}
\begin{center}
\parbox{9cm}{
\includegraphics[width=9cm]{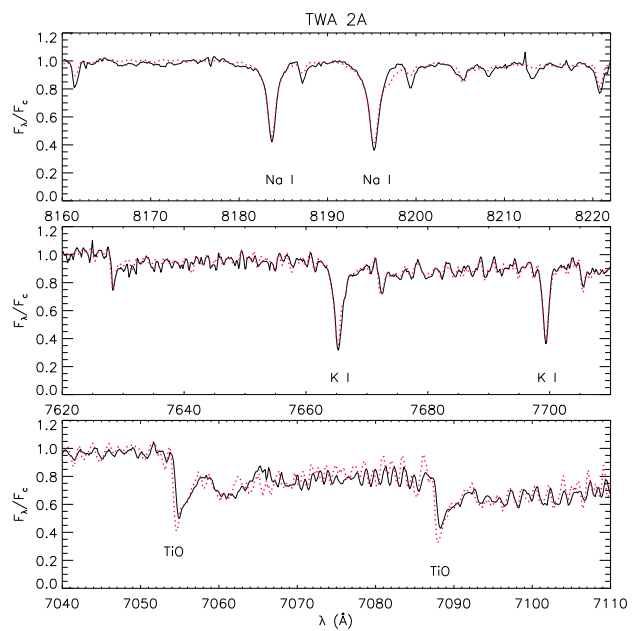}
}
\caption{Observed (solid black lines) and fitted synthetic spectrum (dotted 
red lines) of 
%CD-29\,8887A 
\object{TWA\,2A} in the three spectral regions
chosen to determine the atmospheric parameters with ROTFIT. The most prominent
absorption features are labeled.}
\label{fig:ROTFIT}
\end{center}
\end{figure}

To determinate the surface gravity we selected three additional spectral regions that contain  
the \ion{Na}{i} doublet at $\lambda\approx 8190$\,\AA, the \ion{K}{i} doublet at 
$\lambda\approx 7660-7700$\,\AA, and a spectral segment from 7020 to 7150\,\AA\, 
that contains three TiO molecular bands sensitive to both gravity and temperature.  
In Fig.\,\ref{fig:ROTFIT} we display an example of the result of the fit 
performed by ROTFIT in these three regions. 
As for the $v\sin i$ region, the observed and template spectra were scaled one over the 
other by means of continuum (or pseudo-continuum) windows close to the analysed spectral 
diagnostics. The adopted value of $\log g$ is the weighted average again made  
with the five best templates for each of these three regions. 
For $T_{\rm eff}$ we also included in the average the results of the analysis of the 
spectral segment around 9700\,\AA, which contains several lines whose intensity 
depends on the temperature, but we excluded 
the \ion{Na}{i} doublet, which has a very strong dependency
on $\log g$ that can mask the lower sensibility to $T_{\rm eff}$.

\subsubsection{Results and comparison with the literature}\label{subsubsect:analysis_rotfit_results}

\begin{figure*}
\begin{center}
\parbox{18cm}{
\parbox{9cm}{
\includegraphics[width=9cm]{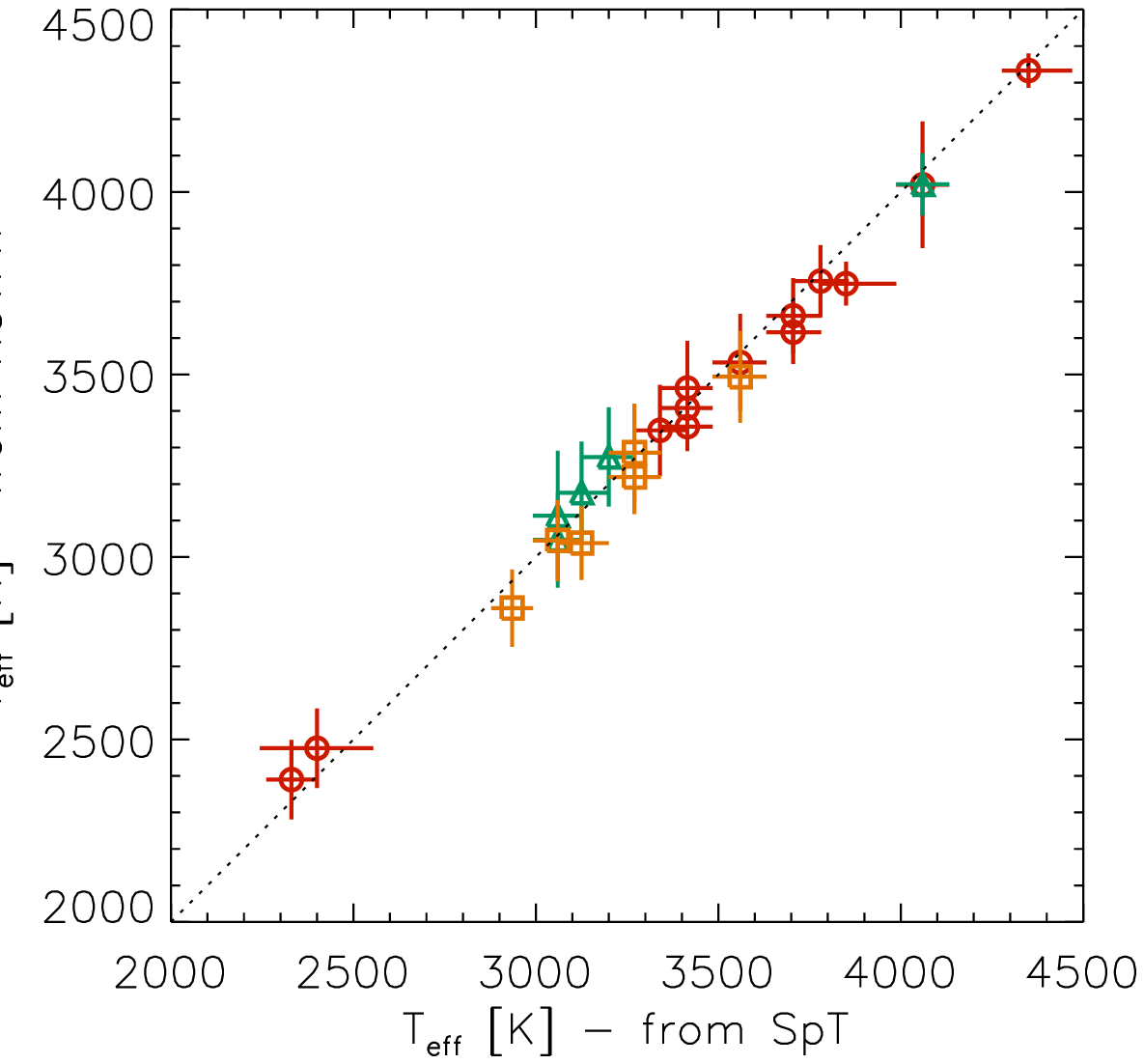}
}
\parbox{9cm}{
\includegraphics[width=9cm]{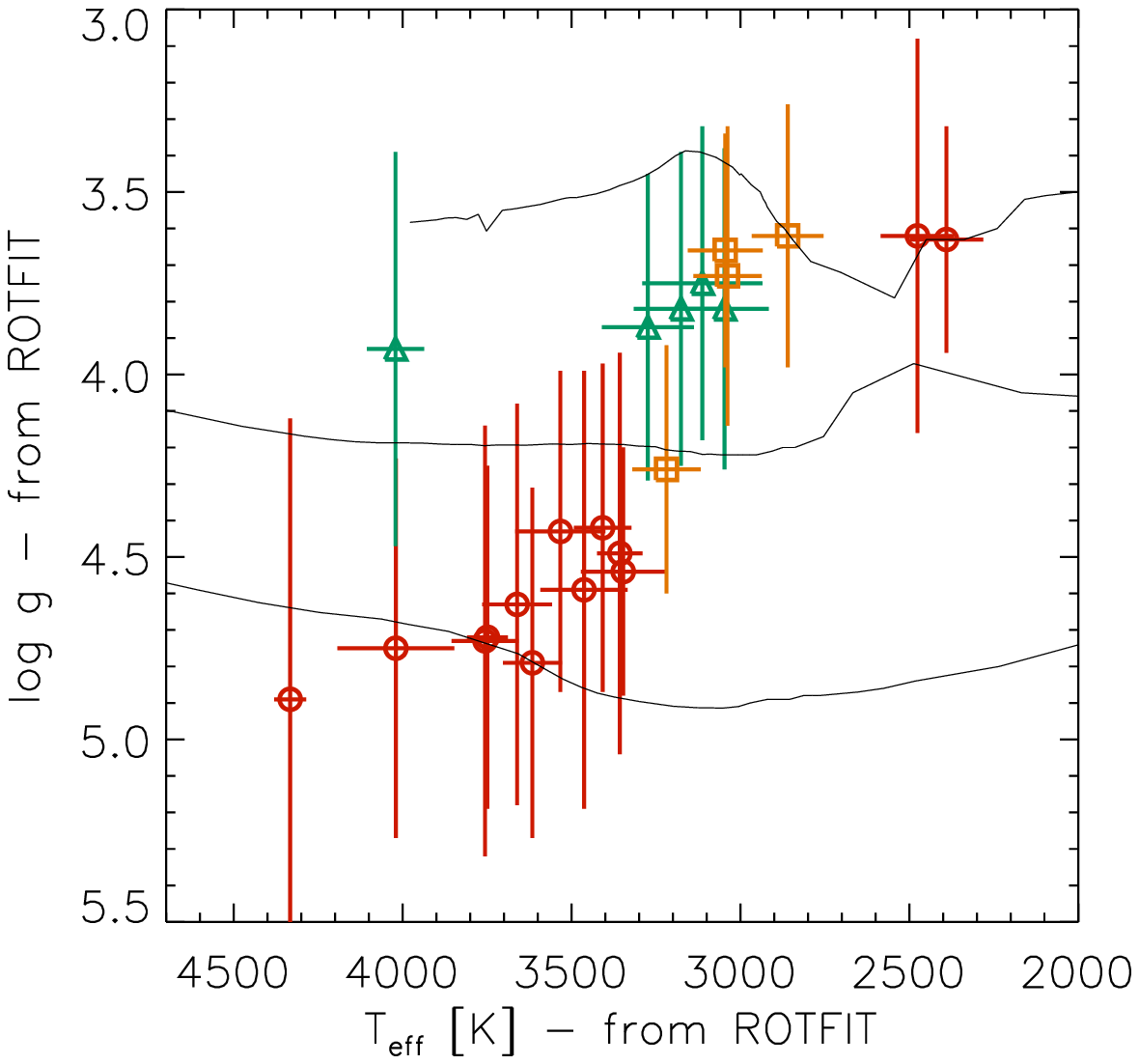}
}
}
\caption{{\it Left panel} -- Comparison between effective temperatures obtained by 
%\protect\cite{Manara13.0} 
\mtr from the SpT with the temperature scale of 
\protect\cite{Luhman03.2} and those derived here with ROTFIT. 
{\it Right panel} -- Surface gravity vs. effective temperature both derived with ROTFIT 
and compared with the predictions of evolutionary PMS models by 
\protect\cite{Baraffe98.1} and \protect\cite{Chabrier00.2},  
for which we show as black lines the 
$1$, $10$, and $100$\,Myr isochrones (from top to bottom).  
Different colors and plotting symbols denote stars from 
the three SFRs: green triangles -- $\sigma$\,Ori, orange squares -- Lupus, red circles --
TWA. Sz\,121 and Sz\,122 are excluded from the right panel because their broad
lines did not allow us to determine the gravity.} 
\label{fig:teff_spt_rotfit}
\end{center}
\end{figure*}

The left panel of Fig.~\ref{fig:teff_spt_rotfit} demonstrates that 
the effective temperatures agree excellently with those of 
%\cite{Manara13.0}. 
\mtr. They obtained the temperatures from the spectral types 
using the conversion presented by \cite{Luhman03.2}, and this empirical temperature scale 
is here confirmed throughout the whole M spectral sequence. 
The derived surface gravities range from $\log{g} \approx 3.6$ to $\approx 4.9$ 
for our targets. 
With the exception of the two coolest objects, 
the two BDs \object{TWA\,26} and \object{TWA\,29}, all stars in the TWA have 
systematically higher gravities than those in Lupus and $\sigma$\,Ori, which is  
qualitatively consistent with the canonical age difference between the three SFRs 
(right panel of Fig.~\ref{fig:teff_spt_rotfit}).  
%consistent with larger age. 
The $1$, $10$, and $100$\,Myr
isochrones of the PMS evolutionary models from 
%\cite{Siess00.1} 
\cite{Baraffe98.1} and \cite{Chabrier00.2} 
are overplotted on the $\log{g}$ vs $T_{\rm eff}$ diagram.  
{\sc `nextgen'} and {\sc `dusty'} models are joined here at $3000$\,K.
The gravities that we derived for the M stars in the TWA are systematically higher 
than predicted by the models for an age of $\sim 10-15$\,Myr, although the effect is
only marginal considering the uncertainties. 
For the two BDs we confirm the similar result of \cite{Mohanty04.2}, who found lower gravities 
than predicted by the \cite{Baraffe98.1} and \cite{Chabrier00.2} models 
for a sample of BDs in the $\approx 11$\,Myr old Upper Sco region \citep[see][for a
comprehensive discussion of the age of Upper Sco]{Pecaut12.0}. 
As a consequence of the low gravity, the BDs appear younger 
in the HR diagram than the M stars located in the same star-forming environment (see 
Fig.~6 of \mtr). 
In Fig.~\ref{fig:teff_spt_rotfit} (right) we have excluded Sz\,122 and Sz\,121 
because their line profiles are dominated
by fast rotation and/or binarity and we could not determine reliable values for their
gravities. One of the remaining stars in Lupus, Sz\,94, has higher gravity than the other
objects in the same SFR. This star is later discarded from our sample due to the
lack of lithium absorption (see Sect.~\ref{subsect:analysis_lithium}). 

Sz\,122 and Sz\,121 are the only two stars from Lupus in our sample with previous
$v \sin{i}$ measurements. Similar to our results, \cite{Dubath96.0} have found very high
rotation rates for them. These stars are likely spectroscopic binaries, therefore 
their $v \sin{i}$ values are uncertain (see also Sect.~\ref{subsect:analysis_rv}).
Rotation velocities have previously been presented for $10$ of the stars in the TWA,
%for two of our Class\,III objects in Lupus and 
and the $v \sin{i}$ values from the literature agree well with our ROTFIT measurements. 
No $v \sin{i}$ are published for our targets in $\sigma$\,Ori. 

Photometric measurements of rotation periods ($P_{\rm rot}$) 
are available for half of our sample \citep{Lawson05.1, Cody10.0} and 
are listed in Table~\ref{tab:photosph_params}. 
We have combined these $P_{\rm rot}$ values with our results on $v \sin{i}$ to compute the
lower limit to the stellar radii, $R \sin{i}$. The values are compatible with the radii
$R_{\rm *,SB}$ determined from $L_{\rm bol}$ and $T_{\rm eff}$ within the errors, which 
consider the uncertainties of $v \sin{i}$ and the $0.2$\,dex uncertainty in the 
bolometric luminosities 
%\citep[see][]{Manara13.0}. 
(see \mtr).

To summarize, the comparison with previous measurements and with predictions
of evolutionary models confirms that we have derived reliable estimates of the 
atmospheric parameters. This was achieved
through a careful selection of the spectral regions in which we carried out
the comparison with the model spectra as described in 
Sect.~\ref{subsubsect:analysis_rotfit_regions}.

\subsection{Stellar radius from the Barnes-Evans relation}\label{subsect:analysis_radius}

\begin{figure}
\begin{center}
\includegraphics[width=9cm]{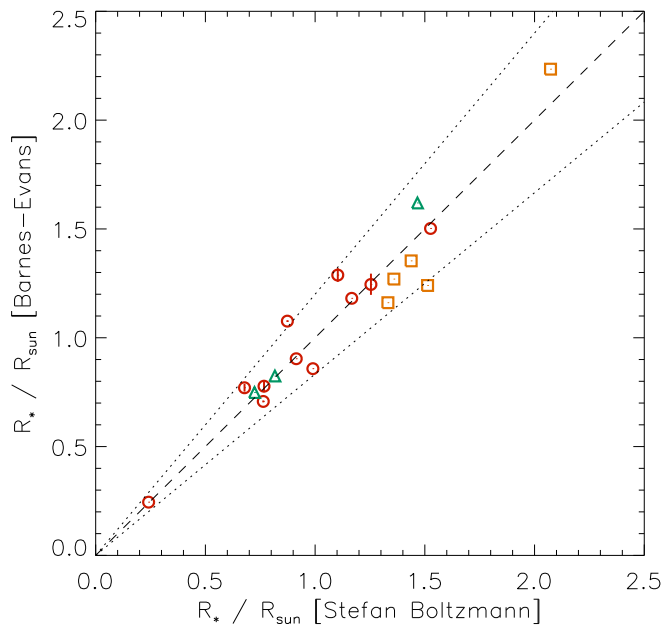}
\caption{Comparison of the stellar radii calculated from the Stefan-Boltzmann law and 
from the Barnes-Evans relation. The $1:1$ line is shown as a dashed line 
and a $20$\,\% difference between the two radii is represented as dotted lines.
Same plotting symbols as in Fig.~\ref{fig:teff_spt_rotfit}.
Not included in this plot are SO\,641, SO\,925, Par-Lup3-1, TWA\,29, and TWA\,14 because
they lack the photometry required for the Barnes-Evans relation (see text
in Sect.~\ref{subsect:analysis_radius}).} 
\label{fig:radius_BE_SB}
\end{center}
\end{figure}

We used published photometry of our targets to determine the stellar radii
from the empirical relation between angular diameter and surface brightness derived by 
\cite{Barnes76.0} for giants that was extended to dwarf stars by \cite{Beuermann99.0}.
Specifically, we applied Eq.~6 of \cite{Beuermann99.0} to the observed $V-K$ colors
after transforming the 2\,MASS $K_s$ magnitudes to the CIT system with the 
transformations given by \cite{Carpenter01.0}.
In Fig.~\ref{fig:radius_BE_SB}, the resulting radii are compared with those obtained from 
the Stefan-Boltzmann law. Four of the five objects that are missing in this figure have no 
measurement of the $V$ magnitude. The fifth one is \object{TWA\,14},  
for which we found a $50$\,\%
difference between the radii computed with the two methods. This can be attributed to 
a problem with the photometry that was previously pointed out by \cite{Zuckerman04.1}. 
For all other stars the two values agree within $20$\,\%, a remarkable fact given that
the calibration of \cite{Beuermann99.0} was based on only eight stars and that their sample
was comprised of young-disk M dwarfs, while ours is composed of PMS stars
with somewhat lower gravity. In Fig.~\ref{fig:radius_BE_SB} the effect of variable
photometry is shown as a vertical bar for stars with more than one measurement in 
the $V$ band; it shows that the spread produced by variability is likely not 
relevant. Finally, some uncertainty may result from extinction. Because our
stars are relatively nearby ($\sim 50-350$\,pc) and have no major circumstellar material
we did not apply a reddening correction to the photometry. 
The comparison of the X-Shooter spectra to spectral templates (see \mtr) shows 
no evidence for reddening within a precision of $\sim 0.5$\,mag. 
Indeed, $A_{\rm V} \sim 0.3$\,mag is sufficient to remove the apparently systematic trend
of the stars in Lupus to be located below the $1:1$ relation in Fig.~\ref{fig:radius_BE_SB}. 

\subsection{Radial velocity}\label{subsect:analysis_rv}

We determined the radial velocity ($RV$) 
of all stars in our sample using the cross-correlation technique.
We calculated the cross-correlation function (CCF) between the observed and
synthetic spectra within ROTFIT to evaluate the velocity shift to be applied 
to the latter. The CCF peak was fitted with a Gaussian to determine more accurately
its center. The standard error of the weighted mean was adopted as the 
measurement error for the atmospheric parameters. 
Despite the poor statistics, it is worth noticing that the 
$RV$ errors tend to be larger for stars with higher $v\sin i$, as expected. 
The barycentric correction was performed with the IDL procedure
{\sc baryvel}. The results of the $RV$ determination are given in 
Table~\ref{tab:photosph_params}.

The radial velocity had previously been measured for all but one of the TWA members 
in our sample. For most stars our results agree well with the values published in 
the literature that have been summarized by \cite{Schneider12.0}. 
Only for two stars (\object{TWA\,7} and \object{TWA\,9A}) the previous and our new 
estimates are 
different by more than $2\,\sigma$. The $RV$ of \object{TWA\,29} has been measured for the
first time here and its value of $7.7 \pm 3.9$\,km/s is consistent with it being
a member of the association. 

\cite{Wichmann99.1} have studied the radial velocities 
in the Lupus star-forming region. 
They distinguished between accreting and non-accreting YSOs in 
the usual way on the basis of the H$\alpha$ equivalent width and evaluated the $RV$ of 
these two samples, classical T Tauri stars (CTTS) and weak-line T Tauri stars (WTTS), 
separately. 
They showed that the radial velocities of the WTTS are
similar to those of the CTTS for the WTTS sample on the clouds 
($\langle RV_{\rm CTTS} \rangle = -0.03 \pm 1.20$\,km/s; 
$\langle RV_{\rm WTTS,on} \rangle = 1.29 \pm 0.87$\,km/s), while
the WTTS outside the clouds clearly have a higher mean radial velocity,  
$\langle RV_{\rm WTTS,off} \rangle = 3.17 \pm 1.39$\,km/s.
Considering the uncertainties, the X-Shooter Class\,III sample in Lupus is compatible
with the two WTTS populations examined by \cite{Wichmann99.1} 
if we exclude two stars with significantly higher values.
The two stars that we excluded from the $RV$ average are Sz\,121 and Sz\,122. 
Sz\,122 is the star with the highest $v\sin i$ ($\approx 150$\,km/s). Its CCF 
shows a double-shaped peak that could be the result of Doppler bumps 
produced by starspots. However, given its very high rotation velocity, 
we think it is more likely the fingerprint of binarity.
Moreover, the RV values given by \cite{Dubath96.0} for Sz\,122 and Sz\,121
are inconsistent with our results, pointing at variability related to a companion. 
More observations are needed to confirm the binary nature of these two stars.

For $\sigma$\,Ori, \cite{Sacco08.2} studied nearly $100$ stars with spectral types K6-M5
and found an $RV$ distribution centered on $+30.93$\,km/s
with a standard deviation of $0.92$\,km/s. The radial velocities we derived for
the X-Shooter sample in $\sigma$\,Ori range from $+29.5$\,km/s to $+35.0$\,km/s 
and agree excellently with the mean cluster value within their uncertainties.

\subsection{Lithium absorption}\label{subsect:analysis_lithium}

The lithium absorption feature at $6708$\,\AA~ is a widely used youth
indicator for low-mass stars because for %$M_* \lsim 0.5\,M_\odot$ 
fully convective stars 
this element is rapidly depleted during the PMS evolution \citep{Bildsten97.0}. 
As already mentioned by 
\mtr, 
%\cite{Manara13.0}, 
we have detected this \ion{Li}{\sc i} line in all stars except for Sz\,94. 
All other properties, for instance the position in the HR diagram, our $RV$ measurement 
and the proper
motion given by \cite{LopezMarti11.0}, are compatible with Sz\,94 being a Lupus member.
However, the absence of lithium in such a young PMS star is difficult to explain, 
and we conclude that this is very likely a foreground object with Lupus-like kinematics
by coincidence. In any case, for this study we discarded Sz\,94 from our sample. 

\begin{figure}
\begin{center}
\includegraphics[width=9cm]{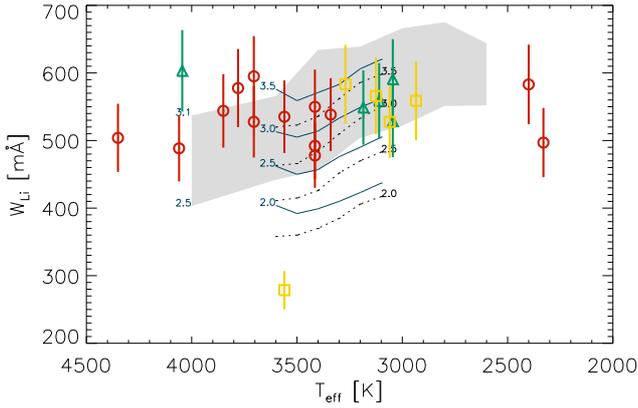}
\caption{Lithium equivalent width versus effective temperature. 
Same plotting symbols as in Fig.~\ref{fig:teff_spt_rotfit}. For clarity 
tiny horizontal offsets are applied to the data for $\sigma$\,Ori and Lupus stars.
Overplotted are curves of growth from 
\protect\cite{ZapateroOsorio02.2} for $\log{g} = 4.5$ and a range of A(Li)$= 2.5-3.1$
as gray shaded and from 
\protect\cite{Palla07.0} for $\log{g} = 4.0$ (black dotted) and $4.5$ (blue solid) 
and four different abundances between A(Li) $= 2.0$ and $3.5$ as labeled.} 
\label{fig:wli_spt}
\end{center}
\end{figure}

The lithium equivalent width ($W_{\rm Li}$) was determined within 
IRAF\footnote{IRAF is distributed by the National Optical Astronomy Observatories, 
which are operated by the Association of Universities for Research in Astronomy, Inc., 
under cooperative agreement with the National Science Foundation.} 
from a by-eye estimate of the local continuum. 
The results are given in Table~\ref{tab:photosph_params}. 
The uncertainties represent the mean and
standard deviations from three measurements carried out on the line, to which we have
added a $10$\,\% systematic error as explained in Sect.~\ref{subsect:data_xshooter}. 
The $W_{\rm Li}$ of the remaining $23$ Class\,III sources 
is shown in Fig.~\ref{fig:wli_spt} as a function of effective temperature. 
There is no evident trend, that is the \ion{Li}{\sc I}\,$\lambda6708$\,\AA~
equivalent width does not depend on effective temperature 
throughout the whole M spectral class. 
After excluding Sz\,94, 
all but one star in our sample have values in the range $\approx 500-600$\,m\AA. 
Only Sz\,122 has much lower $W_{\rm Li}$. We recall that for this star we observed 
very broad lines; it may be a binary. The companion may affect our line measurement in a way
that is impossible to quantify. Qualitatively, the combination of two spectra could
give a very weak lithium absorption if the spectral types of the two components are
very different and one of the two stars has weak lithium. 

In Fig.~\ref{fig:wli_spt} the observed equivalent widths are also compared with curves of
growth calculated by \cite{ZapateroOsorio02.2} and by 
\cite{Palla07.0}. \cite{Palla07.0} provided the lithium abundance, A(Li), in steps of
$0.5$, and we show the curves for $\log{g} = 4.0$ and $4.5$. 
\cite{ZapateroOsorio02.2} covered a wider temperature range. We show their results for 
$\log{g} = 4.5$ in the range A(Li) between $2.5$ and $3.1$ as gray shaded.
The $\log{g}$ values of both calculations are close to the gravities we measured for the 
Class\,III sample (see also Table~\ref{tab:photosph_params}). 
The best agreement between the calculations and our measurements is reached for A(Li) 
$\approx 3.0$, indicating that these stars have retained their primordial
lithium abundance. This is also suggested by the fact that no difference in $W_{\rm Li}$,
and consequently in lithium abundance, is observed in objects belonging to the three SFRs
despite their age difference. Finally, we found good agreement with the lithium measurements 
by \cite{Mentuch08.0} for the $11$ TWA members that we have in common with their study. 

\subsection{Emission lines}\label{subsect:analysis_lines}

The observed fluxes of the major emission lines in the X-Shooter spectra of
our sample have been determined by 
%\cite{Manara13.0}. 
\mtr. These measurements were
obtained by direct integration of the flux above the local continuum. 
Analogous to the case of the lithium absorption line and as explained in 
Sect.~\ref{subsect:data_xshooter}, we assumed a $10$\,\%
systematic error that we added to the statistical measurement errors 
given by \mtr. %\cite{Manara13.0}. 

Generally, accurate line measurements require taking into account the photospheric
absorption component of the lines, for example by subtracting a template spectrum
from the observed spectrum and calculating the emission line flux from the
residual. The template spectrum can be either an observation of an inactive 
star of the same spectral type \citep[see e.g.][]{Frasca94.0,Montes95.3} or a model
spectrum. For late-K and M stars, the only emission lines with a significant
photospheric absorption component are the three Ca\,\ion{\sc ii}\,IRT lines.
For these lines we determined the chromospheric emission line fluxes by applying
the spectral subtraction technique described below that was previously also 
adopted by \mtr. In contrast to that work, we used synthetic
spectra with individual parameters ($T_{\rm eff}$, $\log{g}$ and $v \sin{i}$) for 
each star instead of adopting constant gravity and negligible rotation for all objects. 
For these reasons, we consider
our values for these lines more accurate and adopted the new estimates for 
the subsequent analysis.  
For all other emission lines we made use of the fluxes measured and published by \mtr.
%\cite{Manara13.0}. 
To analyse the \ion{Ca}\,{\sc ii}\,IRT we adopted 
the same BT-Settl synthetic spectra as inactive templates for the spectral subtraction 
that were used to determine the atmospheric parameters.

To check the reliability of the synthetic spectra as inactive templates, we 
compared them with medium-resolution spectra of three low-mass and low-activity stars, 
namely 61~CygA (K5\,V), 61~CygB (K7\,V), and HD1326A (M2\,V), retrieved from the {\em Library 
of Fiber Optic Echelle spectra} of F, G, K, and M field dwarfs \citep{Montes99.0}.
We found that the shape and the residual flux in the cores of 
\ion{Ca}\,{\sc ii}\,IRT lines is 
nearly the same for these low-activity stars and synthetic spectra, which justifies 
the use of the synthetic spectra as inactive templates. 

\begin{figure}
\begin{center}
\includegraphics[width=9cm]{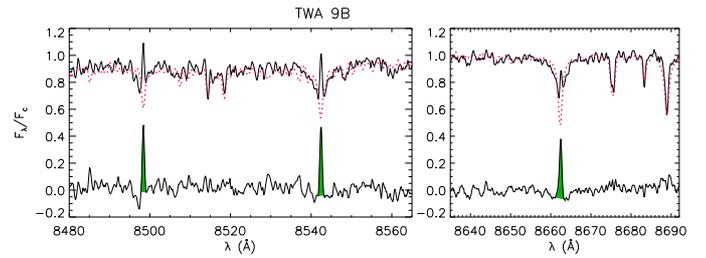}
\caption{Example of the spectral subtraction procedure. The synthetic inactive 
template (dotted red line) is overplotted on the observed continuum-normalized 
spectrum of \object{TWA\,9B} (solid black line). The differences between observed and 
template spectra are plotted at the bottom of each box. The hatched areas in the difference 
spectra represent the excess emissions that have been integrated to obtain the net equivalent 
widths of the \ion{Ca}{\sc ii}\,IRT lines.} 
\label{fig:CaIRT_example}
\end{center}
\end{figure}

To derive the proper surface flux at the
continuum corresponding to the $T_{\rm eff}$ value listed in Table~\ref{tab:photosph_params},
we interpolated the BT-Settl spectra to this temperature. 
The mean continuum flux in the 
range $8570-8640$\,\AA\  was evaluated both in the target spectrum and in the template.
This enabled us to set both spectra on the same scale of continuum-normalized flux
for the subtraction and convert the net equivalent width ($W^{\rm em}_{\rm i}$) 
of each line $i$, measured integrating the residual spectrum, into flux at 
the stellar surface ($F_{\rm surf,i}$) by simply multiplying it with the 
model continuum flux. As an example, we show in Fig.~\ref{fig:CaIRT_example} how 
we applied the spectral subtraction procedure to \object{TWA\,9B}. 

\begin{figure}
\begin{center}
\includegraphics[width=9cm]{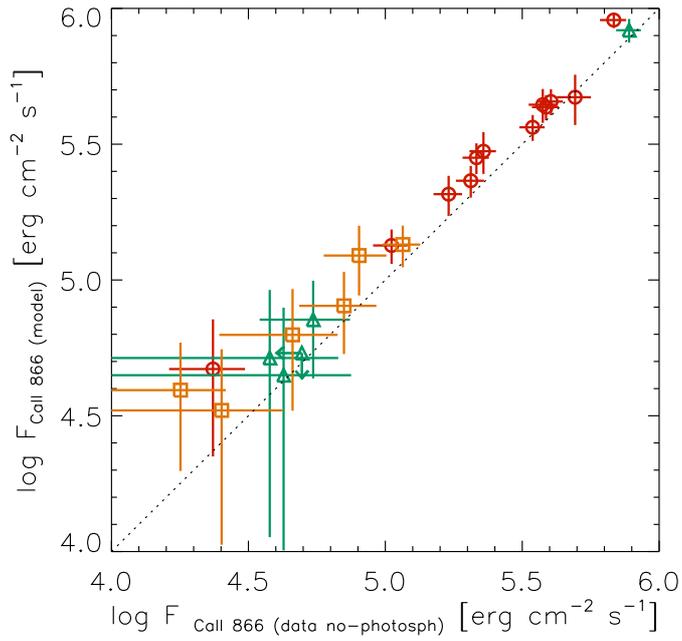}
\caption{Comparison between surface flux of the Ca\,\ion{\sc ii}\,8662\,\AA\, 
line computed in two different ways (see text). Same plotting symbols as in 
Fig.~\ref{fig:teff_spt_rotfit}.}
\label{fig:Fsurf_twomethods}
\end{center}
\end{figure}

Since the observed 
spectra were already flux-calibrated, we analogously derived the flux at Earth 
($f_{\rm obs,i}$) for each line, which was then converted into line luminosity 
($L_{\rm i}$) and line surface flux using the distances and radii 
of the objects listed in Table~\ref{tab:photosph_params}.
We thus obtained for each star two estimates for $F_{\rm surf,i}$, one based on the model
spectrum flux and one based on the observed flux and the dilution factor, $(R_*/d)^2$.
In Fig.~\ref{fig:Fsurf_twomethods} we show the good agreement between the fluxes
derived in the two ways 
for the \ion{Ca}\,{\sc II}$\lambda 8662$\,\AA\, line. This confirms that the
radii and distances we adopted are realistic. 
For consistency with the analysis of the other emission lines where no spectral 
subtraction was necessary, in the remainder of this paper we use for the Ca\,{\sc ii}\,IRT the 
fluxes and derived quantities $f_{\rm obs,i}$ extracted from the observed 
spectrum and not from the model (surface) fluxes. 
These fluxes are summarized in Table~\ref{tab:activity_params}.
In some cases we found no emission above the noise level and the respective 
fields are empty (``...") in Table~\ref{tab:activity_params}. Other cases where 
the equivalent width and flux is consistent with zero within its measurement errors are 
considered as upper limit. 

Following the usual convention, we defined the activity indices as 
\begin{equation}
R^\prime_{\rm i} = \log{(L_{\rm i}/L_{\rm bol})}, 
\label{eq:actindex}
\end{equation}
where $i$ denotes the emission line. The superscript ($^\prime$) indicates that
the photospheric contribution has been subtracted from the line flux for features
with significant photospheric absorption, that is in our case for 
the \ion{Ca}{\sc ii}\,IRT.

\begin{landscape}\begin{table}\begin{center}
\caption{Equivalent widths and fluxes of \ion{Ca}\,{\sc ii}\,IRT.}
\label{tab:activity_params}
\begin{tabular}{l|cccc|cccc|cccc}\hline
\multicolumn{1}{c}{Name} & & \multicolumn{1}{c}{$W_{\rm 8498}$} & \multicolumn{2}{c}{$\log{f_{\rm 8498}}$} & & \multicolumn{1}{c}{$W_{\rm 8542}$} & \multicolumn{2}{c}{$\log{f_{\rm 8542}}$} & & \multicolumn{1}{c}{$W_{\rm 8662}$} & \multicolumn{2}{c}{$\log{f_{\rm 8662}}$} \\
\multicolumn{1}{c}{} & & \multicolumn{1}{c}{{\rm [m\AA]}} & \multicolumn{2}{c}{${\rm [erg/cm^2/s]}$} & & \multicolumn{1}{c}{{\rm [m\AA]}} & \multicolumn{2}{c}{${\rm [erg/cm^2/s]}$} & & \multicolumn{1}{c}{{\rm [m\AA]}} & \multicolumn{2}{c}{${\rm [erg/cm^2/s]}$} \\ \hline
             TWA\,9A & $ $ & $      500 \pm 13$ & $  -13.28$ & $ [-13.33,-13.24]$ & $ $ & $      659 \pm 22$ & $  -13.16$ & $ [-13.21,-13.12]$ & $ $ & $      558 \pm 25$ & $  -13.24$ & $ [-13.29,-13.19]$ \\
             SO\,879 & $ $ & $      618 \pm 16$ & $  -14.23$ & $ [-14.28,-14.19]$ & $ $ & $      806 \pm 18$ & $  -14.11$ & $ [-14.16,-14.07]$ & $ $ & $      696 \pm 18$ & $  -14.18$ & $ [-14.23,-14.14]$ \\
              TWA\,6 & $ $ & $      355 \pm 38$ & $  -13.46$ & $ [-13.53,-13.40]$ & $ $ & $      402 \pm 35$ & $  -13.33$ & $ [-13.39,-13.27]$ & $ $ & $      384 \pm 39$ & $  -13.35$ & $ [-13.41,-13.29]$ \\
             TWA\,25 & $ $ & $      477 \pm 15$ & $  -13.02$ & $ [-13.07,-12.98]$ & $ $ & $      590 \pm 19$ & $  -12.93$ & $ [-12.98,-12.89]$ & $ $ & $      484 \pm 16$ & $  -13.01$ & $ [-13.06,-12.97]$ \\
             TWA\,14 & $ $ & $      397 \pm 22$ & $  -13.93$ & $ [-13.99,-13.89]$ & $ $ & $      586 \pm 23$ & $  -13.68$ & $ [-13.73,-13.64]$ & $ $ & $      485 \pm 24$ & $  -13.77$ & $ [-13.82,-13.72]$ \\
            TWA\,13B & $ $ & $      412 \pm 15$ & $  -13.22$ & $ [-13.27,-13.18]$ & $ $ & $      506 \pm 17$ & $  -13.14$ & $ [-13.18,-13.09]$ & $ $ & $      429 \pm 15$ & $  -13.21$ & $ [-13.26,-13.16]$ \\
            TWA\,13A & $ $ & $      491 \pm 19$ & $  -13.07$ & $ [-13.12,-13.02]$ & $ $ & $      640 \pm 26$ & $  -12.95$ & $ [-13.00,-12.91]$ & $ $ & $      515 \pm 26$ & $  -13.05$ & $ [-13.10,-13.00]$ \\
             TWA\,2A & $ $ & $      317 \pm 18$ & $  -13.00$ & $ [-13.06,-12.96]$ & $ $ & $      366 \pm 16$ & $  -12.94$ & $ [-12.99,-12.90]$ & $ $ & $      295 \pm 19$ & $  -13.03$ & $ [-13.09,-12.99]$ \\
             Sz\,122 & $ $ & $      465 \pm 34$ & $  -14.18$ & $ [-14.24,-14.13]$ & $ $ & $      295 \pm 36$ & $  -14.38$ & $ [-14.45,-14.32]$ & $ $ & $      194 \pm 23$ & $  -14.56$ & $ [-14.63,-14.50]$ \\
             TWA\,9B & $ $ & $      296 \pm 23$ & $  -14.08$ & $ [-14.14,-14.03]$ & $ $ & $      293 \pm 23$ & $  -14.09$ & $ [-14.15,-14.04]$ & $ $ & $      246 \pm 26$ & $  -14.16$ & $ [-14.23,-14.11]$ \\
            TWA\,15B & $ $ & $      524 \pm 23$ & $  -14.08$ & $ [-14.13,-14.04]$ & $ $ & $      619 \pm 22$ & $  -14.01$ & $ [-14.06,-13.97]$ & $ $ & $      474 \pm 22$ & $  -14.12$ & $ [-14.18,-14.08]$ \\
              TWA\,7 & $ $ & $      440 \pm 22$ & $  -13.13$ & $ [-13.18,-13.08]$ & $ $ & $      491 \pm 24$ & $  -13.00$ & $ [-13.05,-12.95]$ & $ $ & $      389 \pm 24$ & $  -13.10$ & $ [-13.16,-13.05]$ \\
            TWA\,15A & $ $ & $      652 \pm 33$ & $  -13.95$ & $ [-14.00,-13.90]$ & $ $ & $      703 \pm 26$ & $  -13.92$ & $ [-13.97,-13.87]$ & $ $ & $      541 \pm 26$ & $  -14.03$ & $ [-14.08,-13.98]$ \\
             Sz\,121 & $ $ & $       65 \pm 25$ & $  -14.94$ & $ [-15.16,-14.80]$ & $ $ & $      109 \pm 43$ & $  -14.72$ & $ [-14.94,-14.57]$ & $ $ & $      250 \pm 59$ & $  -14.35$ & $ [-14.48,-14.25]$ \\
              Sz\,94 & $ $ & $      151 \pm 57$ & $  -14.88$ & $ [-15.09,-14.73]$ & $ $ & $      132 \pm 54$ & $  -14.93$ & $ [-15.17,-14.78]$ & $ $ & $      183 \pm 54$ & $  -14.79$ & $ [-14.96,-14.67]$ \\
             SO\,797 & $ $ & $       71 \pm 46$ & $  -16.25$ & $ [-16.72,-16.03]$ & $ $ & $      332 \pm 71$ & $  -15.58$ & $ [-15.70,-15.49]$ & $ $ & $      144 \pm 50$ & $  -15.95$ & $ [-16.14,-15.81]$ \\
             SO\,641 & $ $ & $      153 \pm 83$ & $  -16.21$ & $ [-16.55,-16.01]$ & $ $ & $     239 \pm 106$ & $  -16.01$ & $ [-16.27,-15.85]$ & $ $ & $      117 \pm 91$ & $  -16.32$ & $ [-16.98,-16.07]$ \\
          Par-Lup3-2 & $ $ & $             ...$ & $     ...$ & $                $ & $<$ & $             203$ & $  -14.83$ & $                $ & $ $ & $      102 \pm 68$ & $  -15.13$ & $ [-15.61,-14.91]$ \\
             SO\,925 & $ $ & $     343 \pm 129$ & $  -15.86$ & $ [-16.08,-15.72]$ & $<$ & $             202$ & $  -16.09$ & $                $ & $<$ & $             148$ & $  -16.23$ & $                $ \\
             SO\,999 & $ $ & $             ...$ & $     ...$ & $                $ & $ $ & $     294 \pm 160$ & $  -15.58$ & $ [-15.93,-15.39]$ & $ $ & $      126 \pm 96$ & $  -15.95$ & $ [-16.58,-15.70]$ \\
             Sz\,107 & $ $ & $      234 \pm 70$ & $  -14.87$ & $ [-15.03,-14.75]$ & $ $ & $      235 \pm 95$ & $  -14.87$ & $ [-15.10,-14.72]$ & $ $ & $      211 \pm 95$ & $  -14.91$ & $ [-15.18,-14.75]$ \\
          Par-Lup3-1 & $<$ & $             107$ & $  -15.93$ & $                $ & $<$ & $             158$ & $  -15.76$ & $                $ & $ $ & $     225 \pm 102$ & $  -15.60$ & $ [-15.88,-15.44]$ \\
             TWA\,26 & $ $ & $             ...$ & $     ...$ & $                $ & $ $ & $             ...$ & $     ...$ & $                $ & $ $ & $             ...$ & $     ...$ & $                $ \\
             TWA\,29 & $ $ & $             ...$ & $     ...$ & $                $ & $ $ & $             ...$ & $     ...$ & $                $ & $ $ & $    1096 \pm 320$ & $  -16.00$ & $ [-16.16,-15.88]$ \\
\hline
\end{tabular}
\end{center}\end{table}\end{landscape}

\section{Magnetic activity}\label{sect:results}

\subsection{Level of H$\alpha$ and X-ray activity}\label{subsect:results_actlevel}

For late-type main-sequence stars the chromospheric and coronal activity 
has long been known to depend on mass, effective temperature, and rotation rate 
\citep[e.g.][]{Wilson66.1, Pallavicini81.1, Noyes84.1}. A correlation of
activity diagnostics with rotation 
rate could never firmly be established for PMS stars and, in fact, \cite{Preibisch05.1}
showed that -- primarily due to their long convective turnover times -- all PMS stars are
in the saturated regime of X-ray activity. The literature is nearly devoid of similar 
studies on chromospheric H$\alpha$ emission of PMS stars because the chromospheric 
contribution of this line is difficult to separate from the effects of accretion and 
non-accreting PMS samples have received less attention. 

\begin{figure*}
\begin{center}
\parbox{18cm}{
\parbox{9cm}{
\includegraphics[width=9cm]{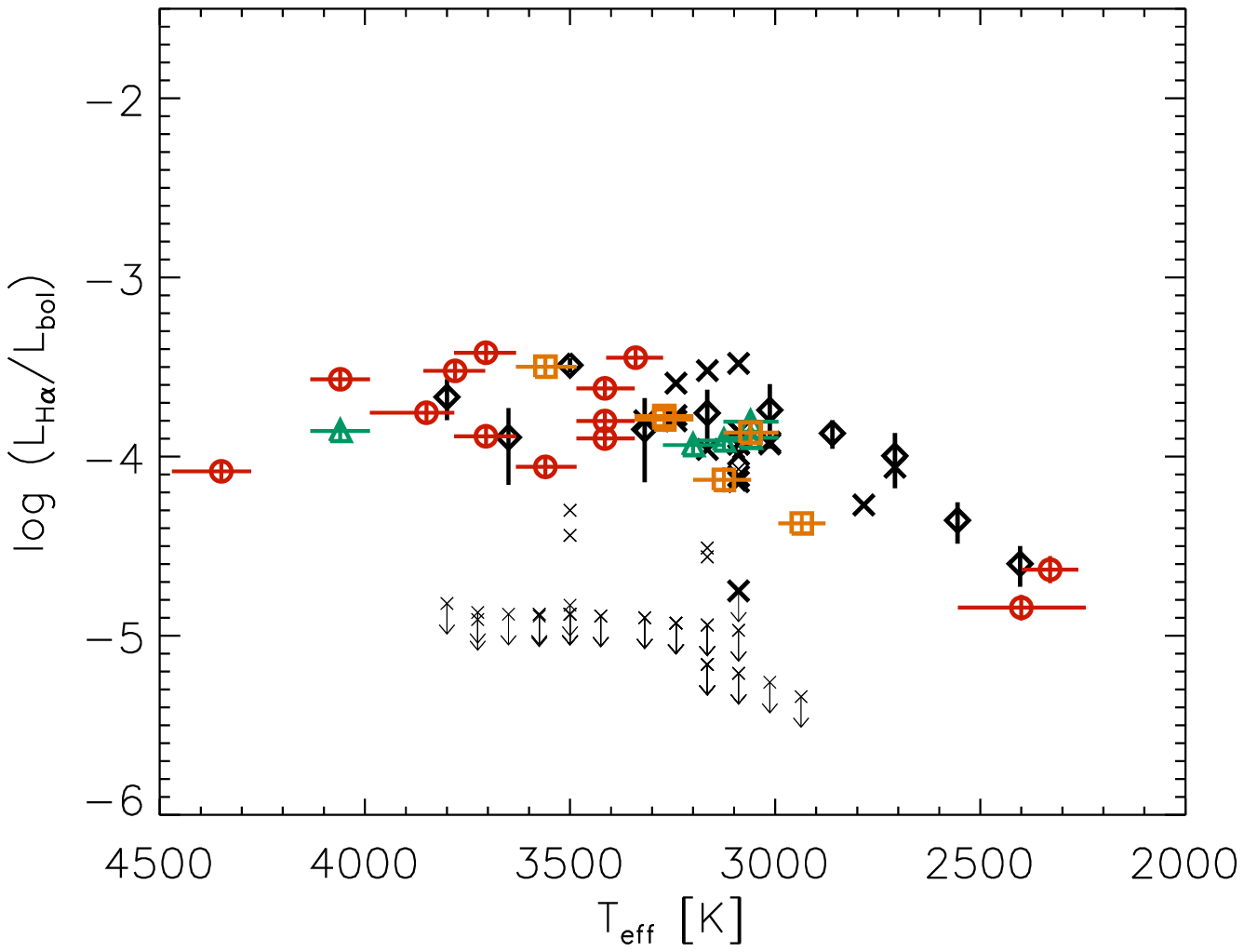}
}
\parbox{9cm}{
\includegraphics[width=9cm]{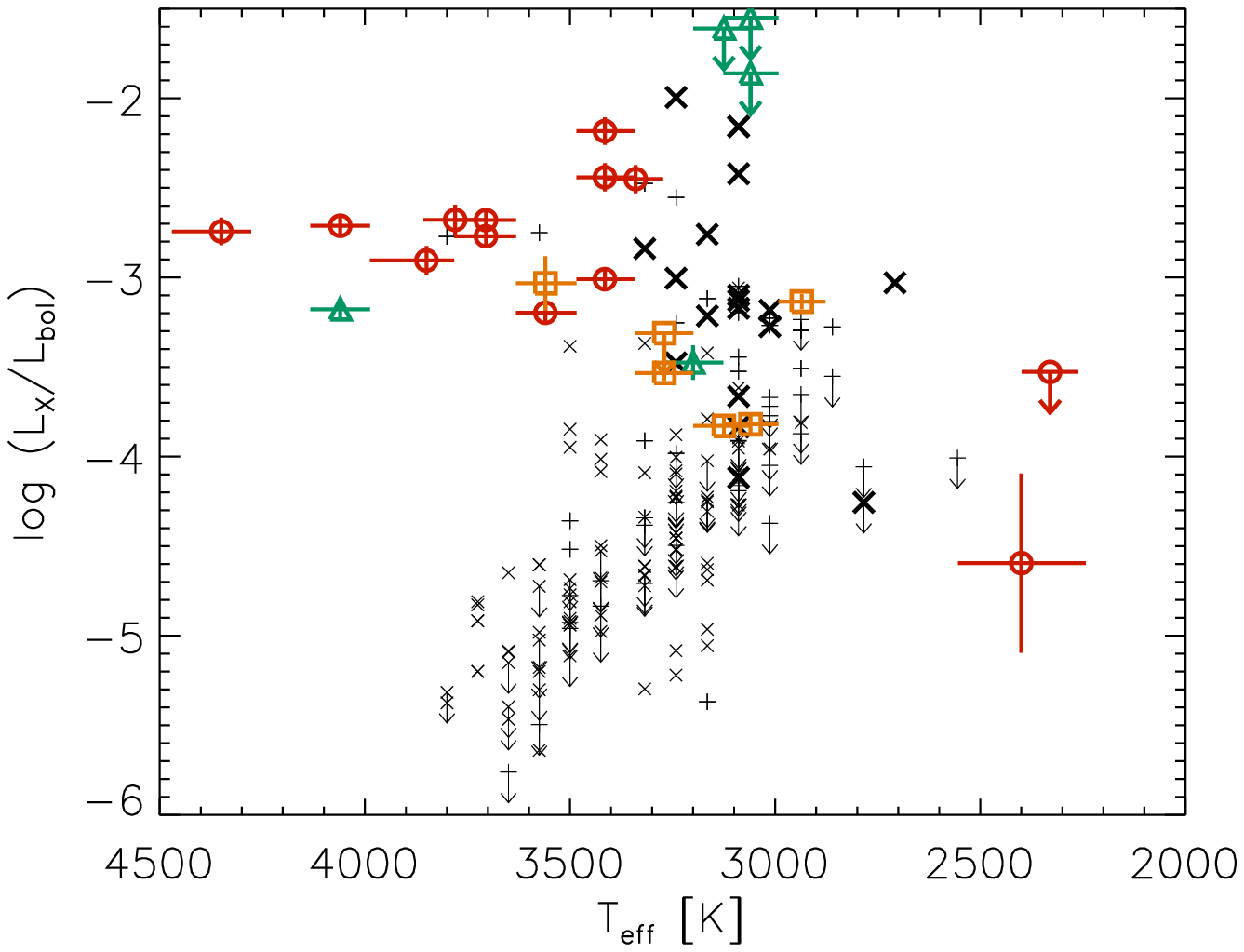}
}
}
\caption{H$\alpha$ and X-ray activity index vs effective temperature for the 
Class\,III sources 
(shown with the same plotting symbols as in Fig.~\ref{fig:teff_spt_rotfit}) 
compared with two samples of field M dwarfs: 
For the $10$-pc sample from \protect\cite{Stelzer13.0} crosses represent stars with
measured rotational velocity (small and large symbols for slow and fast rotators,
respectively, with dividing line at $3$\,km/s) 
and plus symbols are stars without $v \sin{i}$ data. The mean active SDSS 
sample from \protect\cite{Bochanski07.1} is represented by diamonds; 
see Sect.~\ref{subsect:results_actlevel} for details.} 
\label{fig:Rindex_teff}
\end{center}
\end{figure*}

In Fig.~\ref{fig:Rindex_teff} the H$\alpha$ and X-ray activity indices of our
Class\,III sample are shown as a function of $T_{\rm eff}$ % and $v \sin{i}$. 
and are 
compared with two samples of field M dwarfs. Shown as crosses and plus symbols 
is the "$10$-pc sample" of \cite{Stelzer13.0}, which comprises all M dwarfs within 
$10$\,pc of the Sun
from the proper motion survey of \cite{Lepine11.0}. The H$\alpha$ measurements of 
these stars have been compiled from the literature by \cite{Stelzer13.0}, while they 
extracted the X-ray data from public data archives. 
Diamonds in the left panel of Fig.~\ref{fig:Rindex_teff} refer to the work of 
\cite{Bochanski07.1}, who have defined 
template spectra for the M spectral class from several thousands of 
{\em Sloan Digital Sky Survey} (SDSS) spectra. They  
distinguished two subgroups of active and inactive stars on the basis of H$\alpha$
emission and tabulated mean equivalent widths of Balmer lines and the 
$L_{\rm H\alpha}/L_{\rm bol}$ ratio for the active M0...L0 dwarfs. 
These mean values for their `active' sample are the ones shown 
in Fig.~\ref{fig:Rindex_teff}. 

Before comparing field dwarfs and PMS stars, we had to attribute
an effective temperature to the field M dwarfs because the mapping between SpT 
and $T_{\rm eff}$ depends on the evolutionary state of the stars. 
To this end, we combined 
the temperature scales of \cite{Bessell91.1} and \cite{Mohanty03.1},  
which together span the full M spectral sequence.  
Fig.~\ref{fig:Rindex_teff} (left panel) 
shows that for both our PMS sample and the field
star sample from \cite{Bochanski07.1}, $L_{\rm H\alpha}/L_{\rm bol}$ is fairly
constant at a similar level of $\approx -3.5...-4.0$
up to a temperature of $\sim 3000$\,K and drops rapidly towards the
end of the M sequence. 
The plateau for earlier M stars is usually associated with saturation, albeit its
origin is disputed \citep[see e.g.][]{Vilhu87.0, Doyle96.0, Jardine99.0}.
We found 
$\langle \log{(L_{\rm H\alpha}/L_{\rm bol})} \rangle _{\leq \rm M4} = -3.72 \pm 0.21$ 
for the mean H$\alpha$ activity index of the Class\,III sources with spectral type
M4 and earlier, corresponding to $T_{\rm eff} > 3250$\,K. 
The H$\alpha$ activity of the $10$-pc sample shows a much wider spread for given 
$T_{\rm eff}$.
This is because, in contrast to the sample from \cite{Bochanski07.1}, 
the $10$-pc sample comprises many H$\alpha$ inactive stars displayed as downward pointing
arrows in Fig.~\ref{fig:Rindex_teff}. Moreover, this sample exhibits a range of rotation
rates, and these influence the activity level, as can be seen by comparing the large and 
the small crosses that distinguish stars with $v \sin{i} \geq 3$\,km/s from those with 
$v \sin{i} < 3$\,km/s. 
Most of the fast-rotating M dwarfs from the $10$-pc sample have H$\alpha$ activity
similar to that of the PMS sample which consists entirely of fast rotators. 
Therefore, although no rotation rates are given by
\cite{Bochanski07.1}, it is likely that their `active' stars are all rapidly 
rotating. 

An analogous graph for the X-ray activity index is shown in the right panel of 
Fig.~\ref{fig:Rindex_teff}. For our Class\,III sample the dependence of the 
$L_{\rm X}/L_{\rm bol}$ ratio on the $T_{\rm eff}$ is equivalent to that of
H$\alpha$ activity, that is we see a saturation plateau for early-M types that 
declines for cooler objects. The saturation level is higher in X-rays than in H$\alpha$. 
Analogous to the H$\alpha$ emission, 
we computed the mean value for the X-ray activity indices of the Class\,III sources with 
SpT M4 and earlier and found 
$\langle \log{(L_{\rm X}/L_{\rm bol})} \rangle _{\leq \rm M4} = -2.85 \pm 0.36$. 
In the right panel of Fig.~\ref{fig:Rindex_teff} the $10$-pc sample is again overplotted. 
The large divergence of the $L_{\rm X}/L_{\rm bol}$
level between PMS and field dwarfs seen in the early-M stars 
is well-known and commonly attributed to 
rotational evolution and the ensuing decay of activity \cite[e.g.][]{Preibisch05.3}. 
Analogous to the left panel of the same figure,  
as a proxy for the activity state we highlight stars from the $10$-pc sample 
with $v \sin{i} > 3$\,km/s with larger crosses. Stars without known rotation rate 
are marked with a plus symbol.   
The subsample of fast-rotating field M dwarfs
forms the upper boundary and presents $L_{\rm X}/L_{\rm bol}$ ratios similar to
those of the Class\,III stars. 

We have examined the relation between the activity indices and $v \sin{i}$ and
found no dependence, consistent with the observation that our Class\,III stars
have saturated H$\alpha$ and X-ray emission and in agreement with previous results in 
the literature of magnetic activity on PMS stars. 
Only the two BDs have low activity levels considering their high rotational 
velocities. Similarly, \cite{Reiners10.0} found in a sample of late-M stars in the field 
that the fastest rotators show the lowest H$\alpha$ activity.

\subsection{Chromospheric and coronal flux-flux relations}\label{subsect:results_fluxflux}

Several previous works have established power-law relationships between pairs of 
chromospheric emission line fluxes to probe the structure of the outer atmospheres
of late-type stars  
\citep[see e.g.][ henceforth MA11]{MartinezArnaiz11.0}. 
While most samples presented in the literature so far regard field stars, we
examine here a PMS sample. Similar to previous studies, we fitted relations of the type 
\begin{equation}
\log{F_{\rm 1}} = c_1 + c_2 \cdot \log{F_{\rm 2}}, 
\label{eq:fluxflux}
\end{equation}
where $F_{\rm i}$ with $i=1,2$ are the surface fluxes of two lines,  
and the coefficients $c_1$ and $c_2$ are the
free fit parameters. The fits were performed with the least-squares bisector regression 
described by \cite{Isobe90.0}. 
Only stars with detected emission in both diagnostics of the respective 
flux-flux relation were considered. 

The emission lines observed in the X-Shooter spectra comprise a wide spectral
range. We used here the line fluxes for all Balmer lines up to H11, % ($377.0$\,nm), 
Ca\,\ion{\sc ii}\,H\&K, and the two helium lines (He\,\ion{\sc i}\,$\lambda$5876\,\AA\, and 
He\,\ion{\sc i}\,$\lambda$10830\,\AA) listed by \mtr %\cite{Manara13.0} 
and our results for the Ca\,\ion{\sc ii}\,IRT derived as described in 
Sect.~\ref{subsect:analysis_lines}. 
These lines are produced in different layers of the 
outer atmosphere, from the lower chromosphere to the transition region, 
as explained in Sect.~\ref{sect:intro}. 
We also extend our investigation to a comparison between chromospheric and coronal
emission using our compilation of X-ray fluxes.

\begin{figure*}
\begin{center}
\parbox{18cm}{
\parbox{6cm}{
\includegraphics[width=6cm]{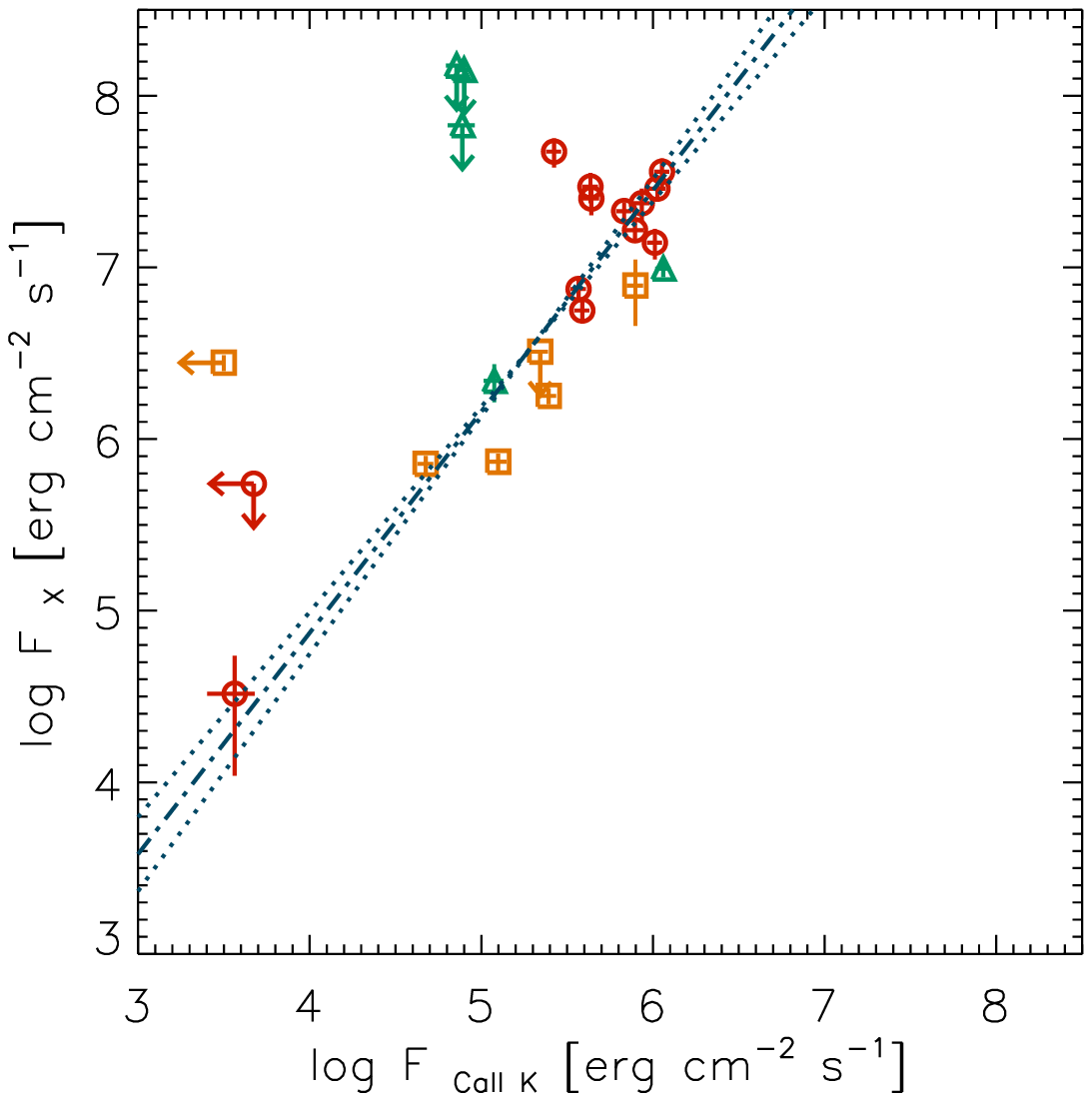}
}
\parbox{6cm}{
\includegraphics[width=6cm]{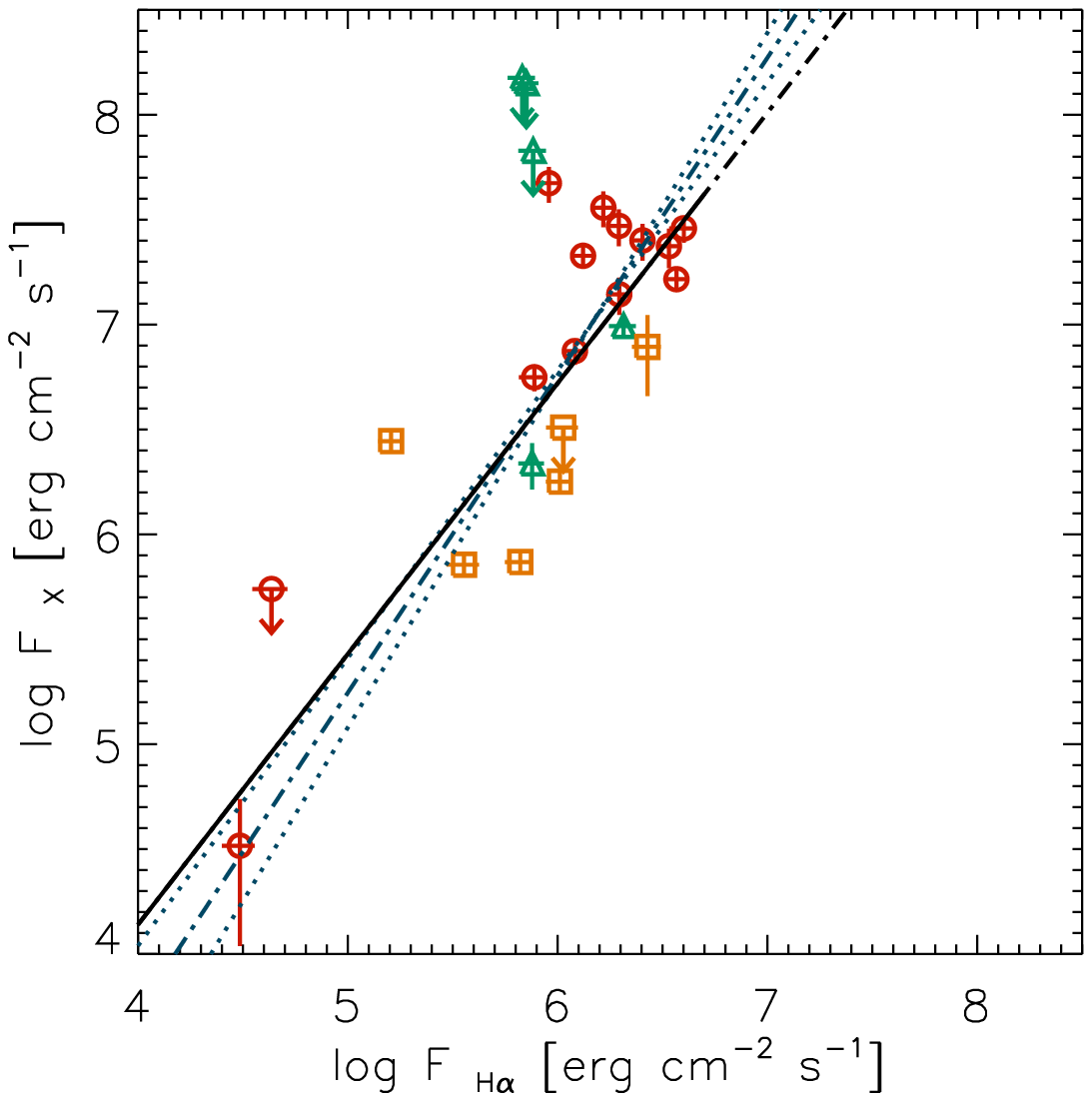}
}
\parbox{6cm}{
\includegraphics[width=6cm]{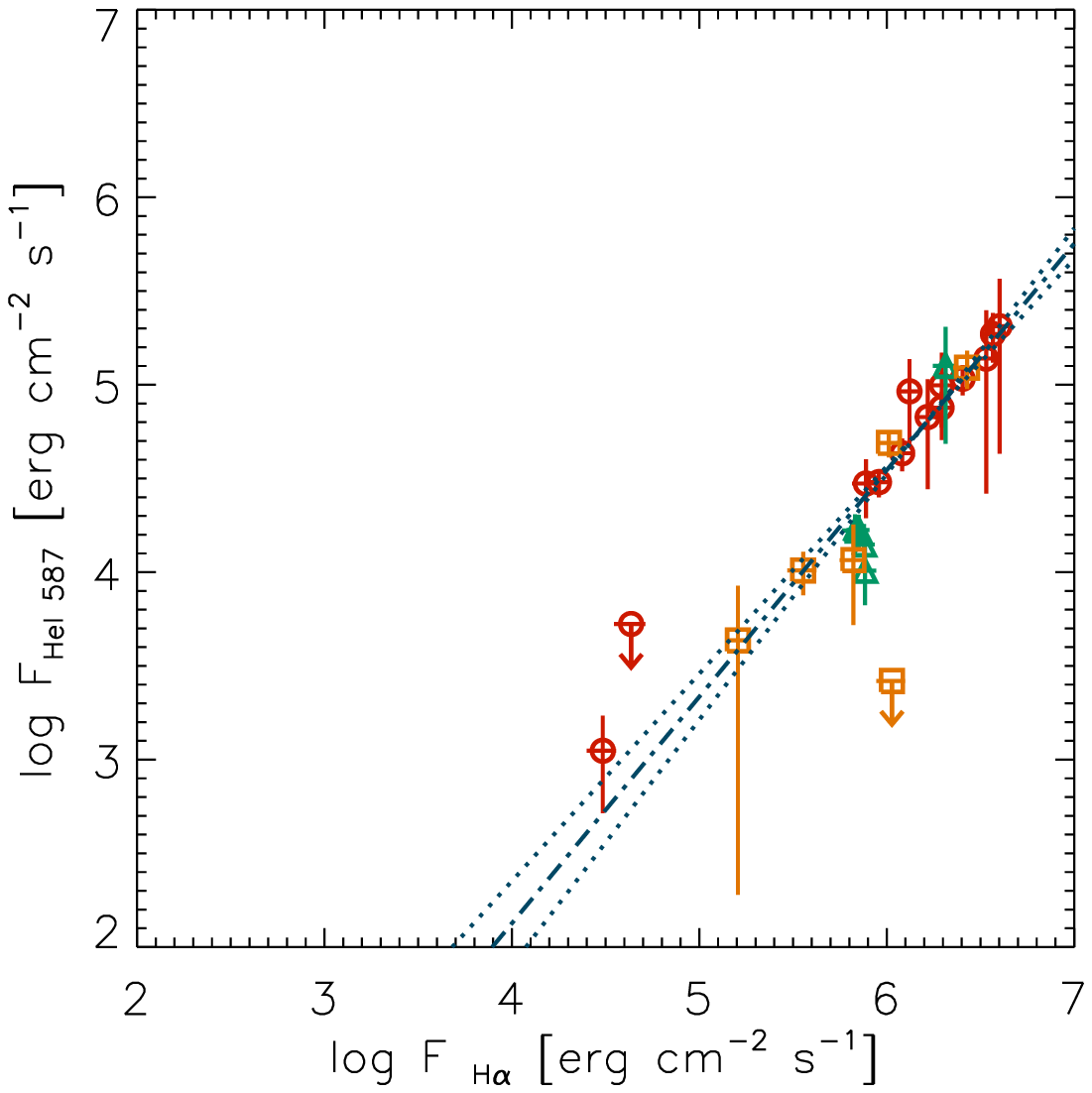}
}
}
\caption{Flux-flux relations for some of the most often used indicators of chromospheric
and coronal activity in M stars 
(H$\alpha$, \ion{Ca}\,{\sc ii}\,K, X-rays and \ion{He}\,{\sc i}$\lambda\,5876$\,\AA). 
Same plotting symbols as in Fig.~\ref{fig:teff_spt_rotfit}. 
The results of a linear
regression fit to all detections and its standard deviation 
are shown as dash-dotted and dotted lines, respectively. For $F_{\rm X}$ vs.
$F_{\rm H\alpha}$ we also show 
the relation derived in the same way for a sample of
field M dwarfs as solid line \protect\citep[see][]{Stelzer12.0}.}
\label{fig:fluxflux_special}
\end{center}
\end{figure*}

\begin{figure*}
\begin{center}
\parbox{18cm}{
\parbox{6cm}{
%CaK_Ha_weights0.ps
\includegraphics[width=6cm]{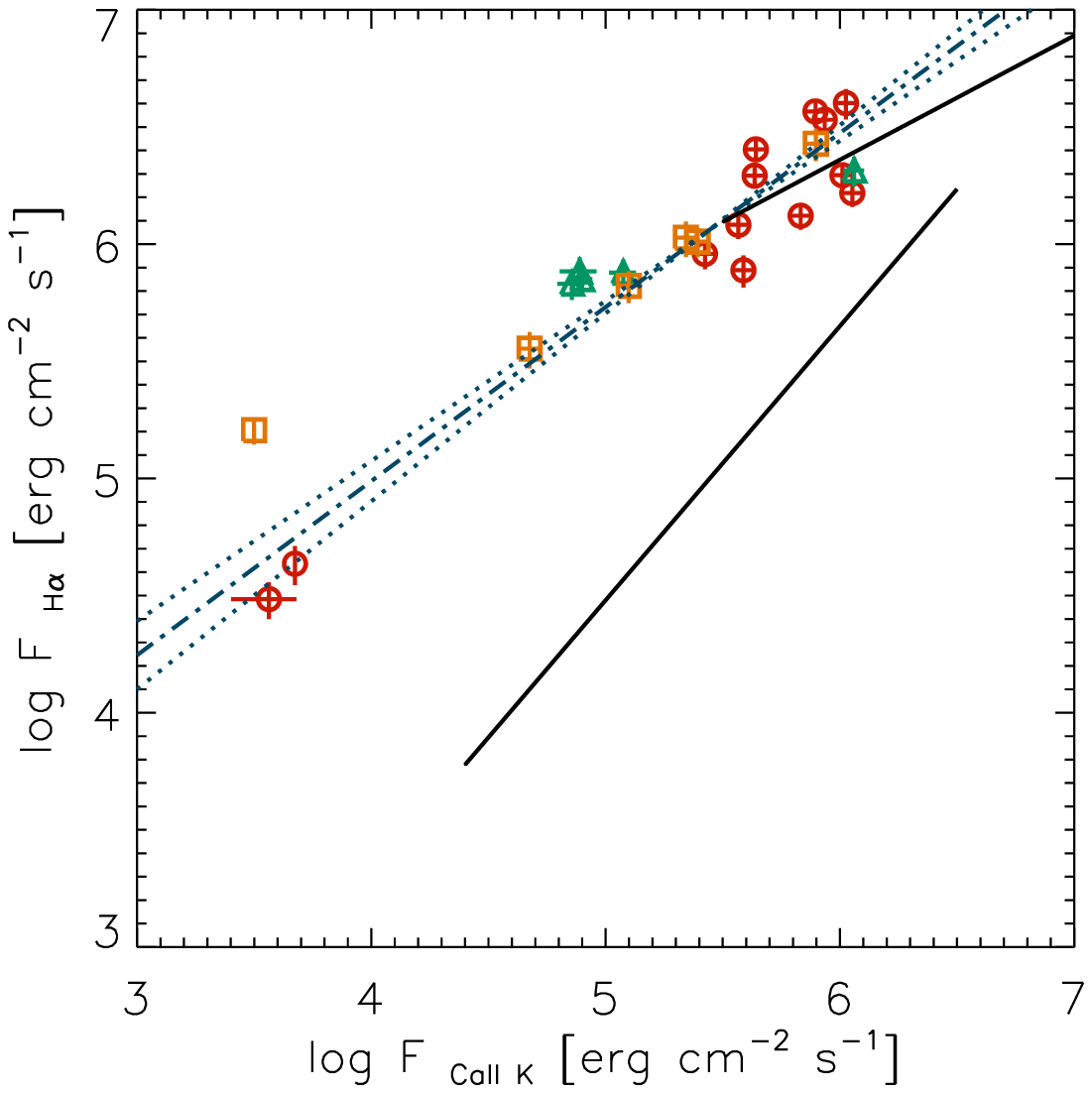}
}
\parbox{6cm}{
%Ca849_Ha_weights0
\includegraphics[width=6cm]{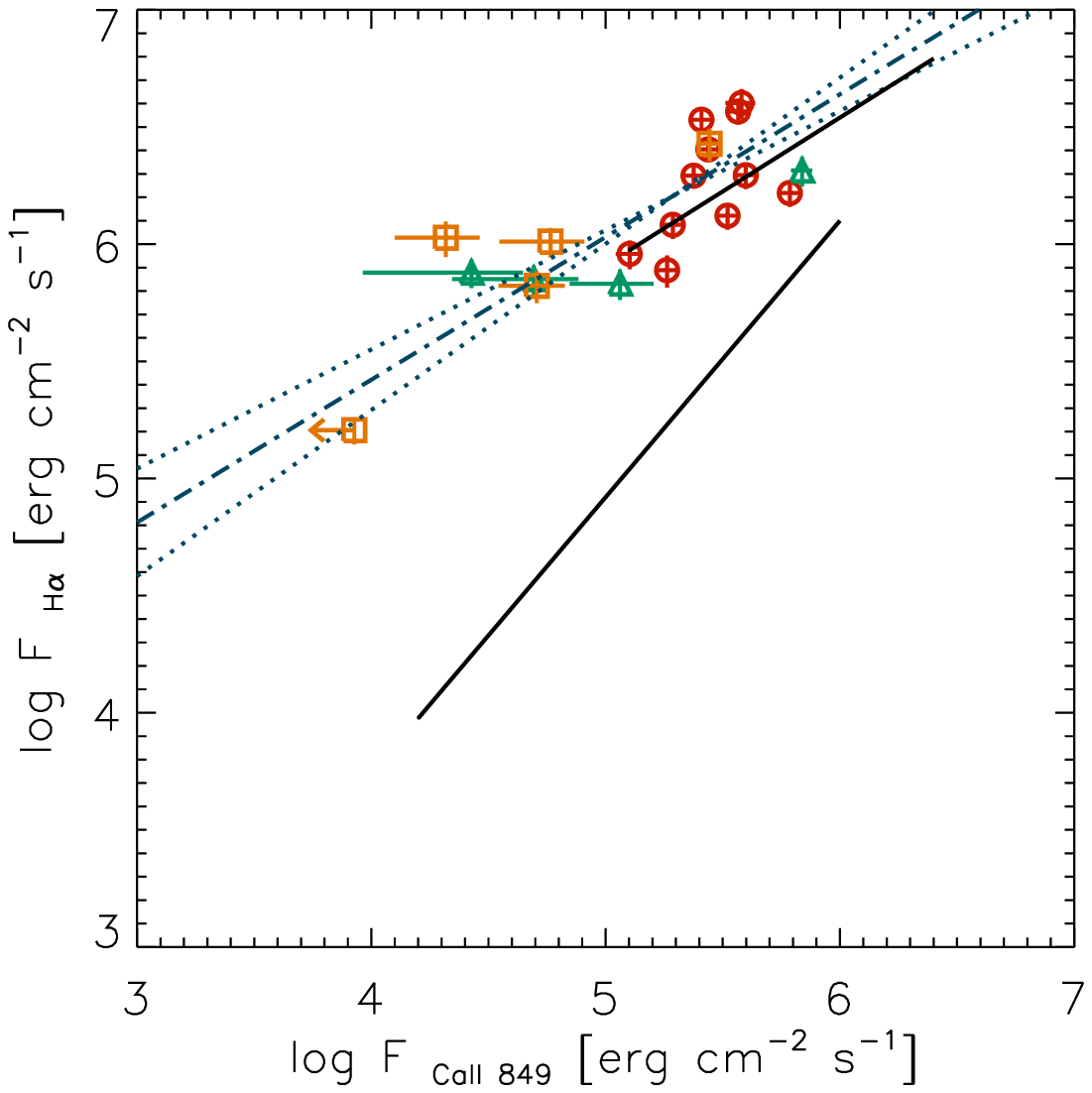}
}
\parbox{6cm}{
%Ca849_X_weights0
\includegraphics[width=6cm]{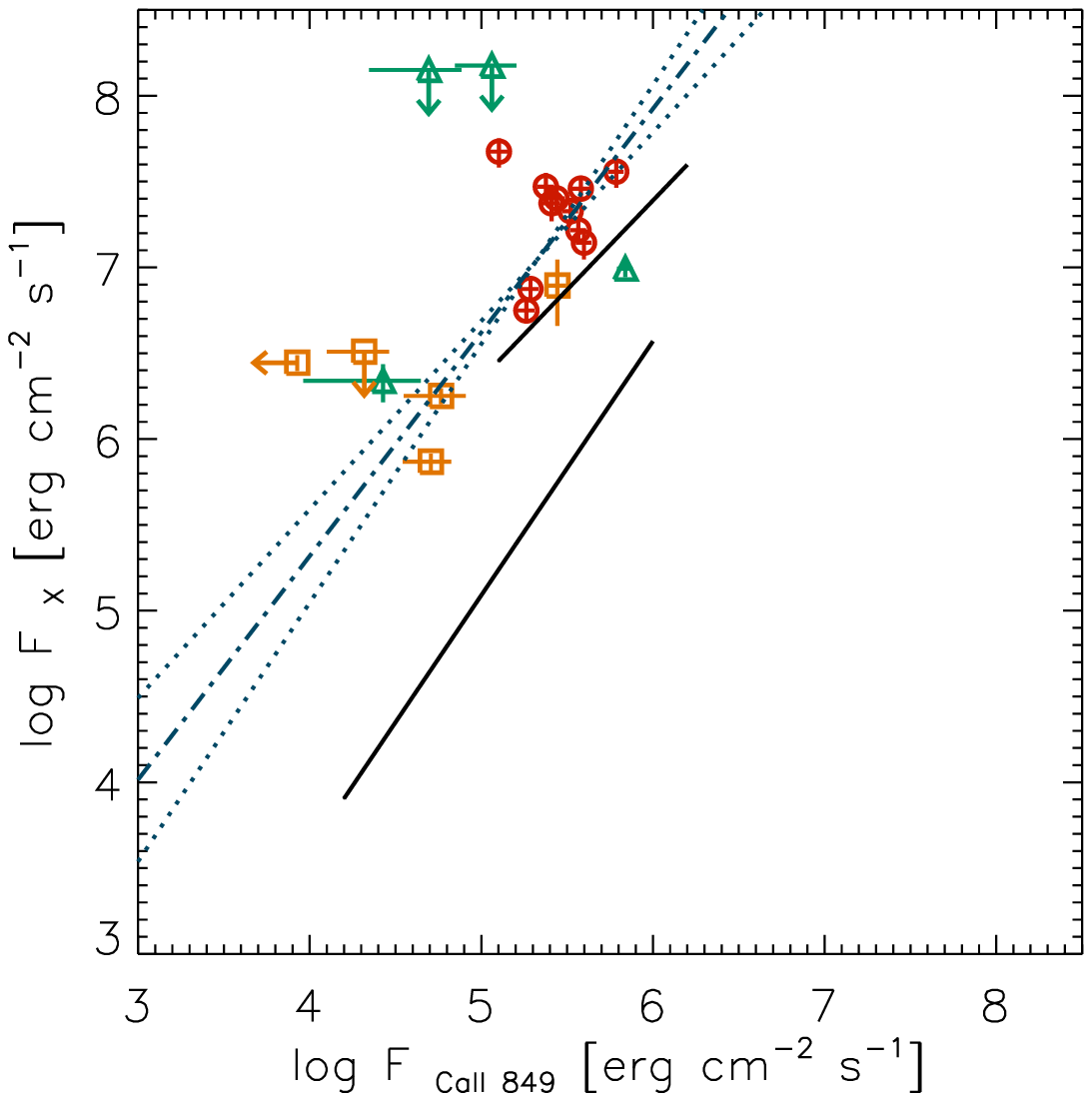}
}
}
\caption{Flux-flux relations for which MA11 identified two distinct populations in a 
sample of field stars (shown as solid lines). Our sample of Class\,III stars 
is shown with the same plotting symbols as in Fig.~\ref{fig:teff_spt_rotfit} and 
is located on the upper relation. It extends this `active branch' to lower fluxes. 
Linear regressions and 
standard deviation for the Class\,III detections are shown as dash-dotted and dotted lines 
as in Fig.~\ref{fig:fluxflux_special}.}
\label{fig:fluxflux_active}
\end{center}
\end{figure*}

The results of the linear regressions of all examined flux pairs are given in
Table~\ref{tab:fluxflux}. Some of the more notable correlations are shown in 
Figs.~\ref{fig:fluxflux_special} and~\ref{fig:fluxflux_active}. 
In these two figures the fitted lines are overplotted (dash-dotted and dotted
for the uncertainties).  
Where available, we also show the regressions published previously for different
samples of late-type stars (solid lines). 
\begin{table}
\begin{center}
\caption{Linear fit coefficients for flux-flux relationships of type
$\log{F_1} = c_1 + c_2 \cdot \log{F_2}$. `$N_*$' is the number of stars
detected in both lines.} 
\label{tab:fluxflux}
\begin{tabular}{ccccc} \\ \hline
Line\,1    & Line\,2    & $N_*$ & $c_1$ & $c_2$ \\
\hline
  H$\beta$ & H$\alpha$   & $ 23 $ & $ -1.71 \pm 0.56 $ & $  1.19 \pm 0.09 $ \\
 H$\gamma$ & H$\alpha$   & $ 23 $ & $ -1.96 \pm 0.70 $ & $  1.18 \pm 0.11 $ \\
 H$\delta$ & H$\alpha$   & $ 23 $ & $ -2.43 \pm 0.97 $ & $  1.23 \pm 0.16 $ \\
        H8 & H$\alpha$   & $ 21 $ & $ -2.37 \pm 0.94 $ & $  1.19 \pm 0.15 $ \\
        H9 & H$\alpha$   & $ 20 $ & $ -4.81 \pm 1.03 $ & $  1.56 \pm 0.17 $ \\
       H10 & H$\alpha$   & $ 20 $ & $ -5.31 \pm 1.06 $ & $  1.62 \pm 0.18 $ \\
       H11 & H$\alpha$   & $ 20 $ & $ -5.42 \pm 1.01 $ & $  1.62 \pm 0.17 $ \\
 H$\gamma$ &  H$\beta$   & $ 23 $ & $ -0.25 \pm 0.18 $ & $  1.00 \pm 0.03 $ \\
 H$\delta$ &  H$\beta$   & $ 23 $ & $ -0.66 \pm 0.33 $ & $  1.04 \pm 0.06 $ \\
 He\,I\,587 & H$\alpha$  & $ 23 $ & $ -2.77 \pm 0.69 $ & $  1.22 \pm 0.11 $ \\
 He\,I\,1083 & H$\alpha$ & $  6 $ & $ -2.94 \pm 2.44 $ & $  1.32 \pm 0.38 $ \\
 H$\alpha$ &   Ca\,II\,K & $ 23 $ & $ +2.35 \pm 0.45 $ & $  0.69 \pm 0.08 $ \\
 H$\alpha$ &   Ca\,II849 & $ 19 $ & $ +2.95 \pm 0.52 $ & $  0.62 \pm 0.10 $ \\
 H$\alpha$ &   Ca\,II854 & $ 19 $ & $ +2.72 \pm 0.51 $ & $  0.65 \pm 0.10 $ \\
 H$\alpha$ &   Ca\,II866 & $ 22 $ & $ +1.31 \pm 0.90 $ & $  0.92 \pm 0.17 $ \\
 He\,I\,587 &  Ca\,II\,K & $ 23 $ & $ +0.08 \pm 0.40 $ & $  0.83 \pm 0.07 $ \\
  Ca\,II\,K &  Ca\,II849 & $ 19 $ & $ +0.89 \pm 0.76 $ & $  0.90 \pm 0.14 $ \\
 He\,I\,587 &  Ca\,II849 & $ 19 $ & $ -0.10 \pm 0.59 $ & $  0.92 \pm 0.11 $ \\
Ca\,II\,849 &  Ca\,II854 & $ 18 $ & $ -0.58 \pm 0.59 $ & $  1.09 \pm 0.10 $ \\
Ca\,II\,849 &  Ca\,II866 & $ 18 $ & $ -0.68 \pm 0.71 $ & $  1.12 \pm 0.13 $ \\
     Xrays & H$\alpha$   & $ 19 $ & $ -2.27 \pm 0.83 $ & $  1.50 \pm 0.14 $ \\
     Xrays &    CaIIK    & $ 19 $ & $ +1.04 \pm 1.01 $ & $  1.06 \pm 0.17 $ \\
     Xrays &   CaII\,849 & $ 16 $ & $ +0.21 \pm 1.09 $ & $  1.28 \pm 0.20 $ \\
    He\,I\,587 &  Xrays  & $ 17 $ & $ +1.65 \pm 0.84 $ & $  1.10 \pm 0.17 $ \\
\hline
\end{tabular}
\end{center}
\end{table}

For Fig.~\ref{fig:fluxflux_special} we selected
correlations between some of the most widely used and most easily accessible indicators 
of magnetic activity (X-rays, H$\alpha$, \ion{Ca}\,{\sc ii}\,K, 
\ion{He}\,{\sc i}\,$\lambda5876$\,\AA).
We compared our results with the relations found in the literature. In particular, 
the correlation between X-ray and H$\alpha$ flux has recently been studied by
MA11 and by \cite{Stelzer12.0}. MA11 considered a sample of nearly $300$ single dwarf
stars with spectral types from F to mid-M, while \cite{Stelzer12.0} have reanalyzed
flux-flux relations for the subsample of M stars from MA11 and combined it with late-M
stars. The slope of the $F_{\rm X}$ vs. $F_{\rm H\alpha}$ relation for our Class\,III sample 
is statistically consistent with those of both these studies 
(see Fig.~\ref{fig:fluxflux_special} middle). 
Moreover, all flux-flux combinations relating the two calcium lines (not shown graphically)
agree excellently with the results of MA11. 

In Fig.~\ref{fig:fluxflux_active} we present the correlations for which MA11
have identified two branches, a main one defined by field stars in a wide range of
spectral types (FGKM) and a second one populated by a subsample of M field dwarfs. In the
study of MA11 this latter group apparently shows higher H$\alpha$ and X-ray flux for given 
\ion{Ca}\,{\sc ii}\,K and \ion{Ca}\,{\sc ii}$\lambda\,8498$\,\AA\, emission. 
The linear regressions computed by
MA11 for the two branches are overplotted as solid black lines in 
Fig.~\ref{fig:fluxflux_active} in the range of fluxes covered by their sample. 
Evidently, our Class\,III sample closely follows the upper `active' branch defined
by MA11, and it allows us to extend the correlations of H$\alpha$ vs. \ion{Ca}\,{\sc ii}\,K,
H$\alpha$ vs. \ion{Ca}\,{\sc ii}$\lambda\,8498$\,\AA, 
and X-rays vs. \ion{Ca}\,{\sc ii}$\lambda\,8498$\,\AA\,  
by $1-2$ dex towards lower fluxes. 
None of the Class\,III objects is located on the main lower branch occupied by 
the majority of field stars according to MA11. 

\begin{figure}
\begin{center}
%logFxFlines_teff
\includegraphics[width=9cm]{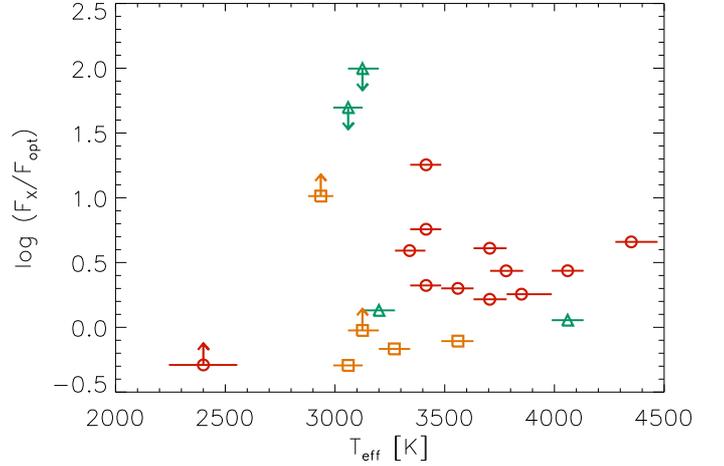}
\caption{Logarithm of the ratio between coronal X-ray flux and chromospheric 
optical flux versus effective temperature. The chromospheric optical flux is
defined as the sum of the fluxes of all strong emission lines in the 
X-Shooter spectra. Same plotting symbols as in Fig.~\ref{fig:teff_spt_rotfit}.} 
\label{fig:corona_vs_chromo}
\end{center}
\end{figure}

Finally, to compare the chromospheric and the coronal radiative output we summed all
optical emission lines studied in this paper, that is the Balmer series up to H11, 
the \ion{He}{\sc i} lines at $5876$\,\AA\, and at $10830$\,\AA,
the three lines of the \ion{Ca}\,{\sc ii}\,IRT, and 
the \ion{Ca}\,{\sc ii}\,K line. We multiplied the flux of
the \ion{Ca}\,{\sc ii}\,K line by a factor of 
two to approximate the \ion{Ca}\,{\sc ii}\,H emission that was omitted from the observed 
line list because of its blend with H$\epsilon$. The contribution of
H$\epsilon$ was approximated as the mean of the observed fluxes of H$\delta$ and H8. The 
summed chromospheric flux in optical emission lines ($F_{\rm opt}$) was 
considered an upper limit if at least one of the examined lines was not detected. 
The ratios between $F_{\rm X}$ and $F_{\rm opt}$ are shown in Fig.~\ref{fig:corona_vs_chromo}. 
Stars in which both the X-ray emission and the chromospheric optical 
emission is an upper limit are not included in Fig.~\ref{fig:corona_vs_chromo} because their 
corona-to-chromosphere flux ratio is completely unconstrained. In most stars the emission 
from the corona dominates that from the visible chromosphere. The difference is typically a 
factor $2$ to $5$, and there is a possible hint for systematically higher 
$F_{\rm X}/F_{\rm opt}$ in TWA members than in those of the younger objects 
from Lupus and $\sigma$\,Ori. 

The full chromospheric radiation budget comprises emission in the near-ultraviolet (NUV)
and far-ultraviolet (FUV). Very few records of ultraviolet emission from Class\,III sources 
are found in the literature. The {\it Hubble Space Telescope} (HST) study of \cite{Yang12.0} 
and the {\it Galaxy Evolution Explorer} (GALEX) study of \cite{Stelzer13.0} have a few stars 
in common with our X-Shooter sample. We list the UV fluxes measured at Earth for these stars 
in 
Table~\ref{tab:chromo_fluxes} 
together with the sum of the fluxes emitted in the optical
emission lines ($f_{\rm opt}$). The HST fluxes refer to the $1250-1700$\,\AA\, band, and
typically about half of them are emitted in the five strongest emission lines 
\citep[see Table~4 in][]{Yang12.0}. The GALEX FUV and NUV bands comprise $1344-1786$\,\AA\, 
and $1771-2831$\,\AA, respectively. The photospheric contribution to the broad-band FUV
and NUV fluxes was subtracted by modeling the individual SEDs 
\citep[see][]{Stelzer13.0}. No information on the distribution of the flux on
continuum and emission lines is available for the GALEX observations. 
While a detailed comparison of optical and UV chromospheric emissions is impeded by the 
small sample size, it seems that the optical contribution dominates. 

\begin{figure*}
\begin{center}
\parbox{18cm}{
\parbox{9cm}{
%HaHb_Teff
\includegraphics[width=9cm]{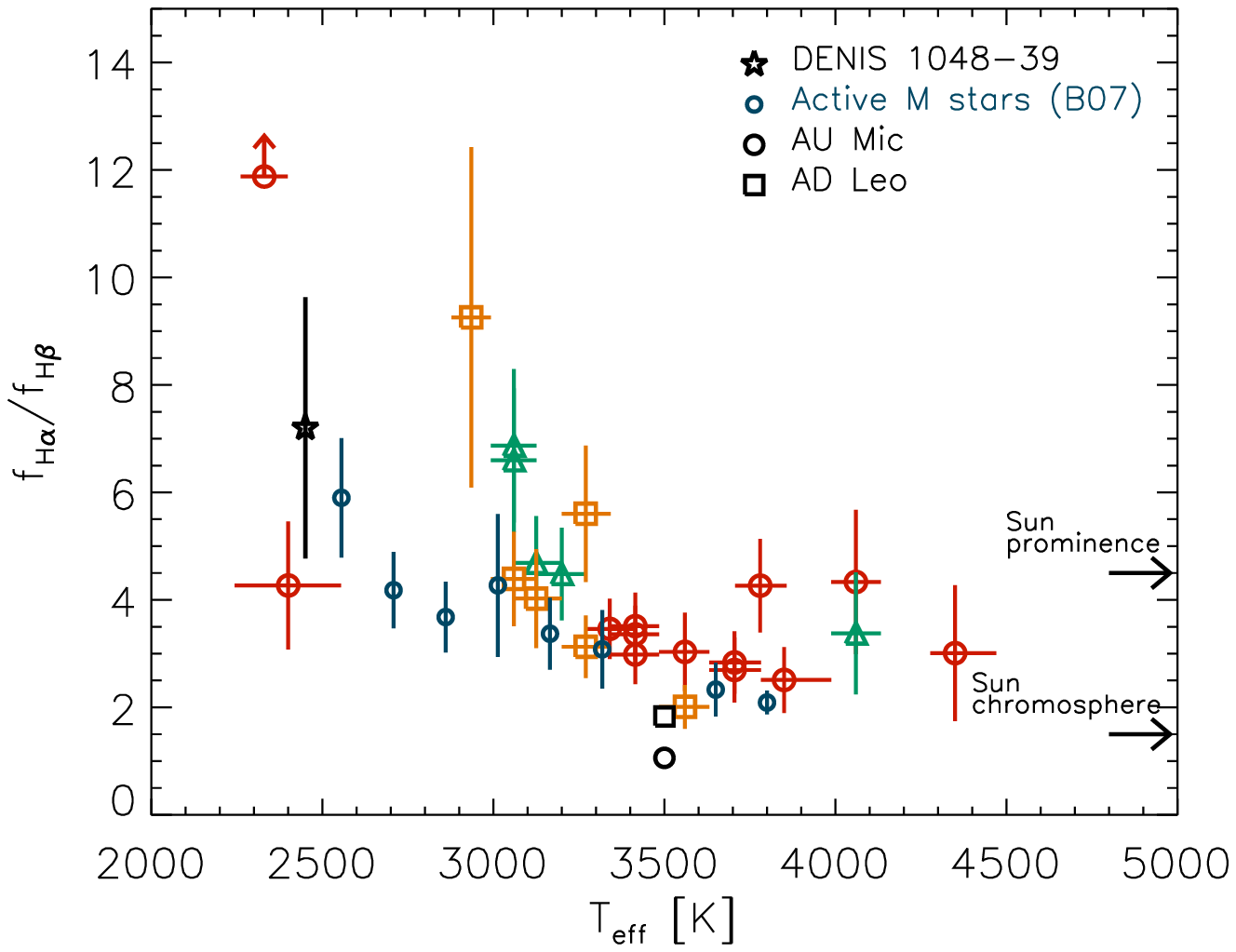}
}
\parbox{9cm}{
%CaKHa_Teff
\includegraphics[width=9cm]{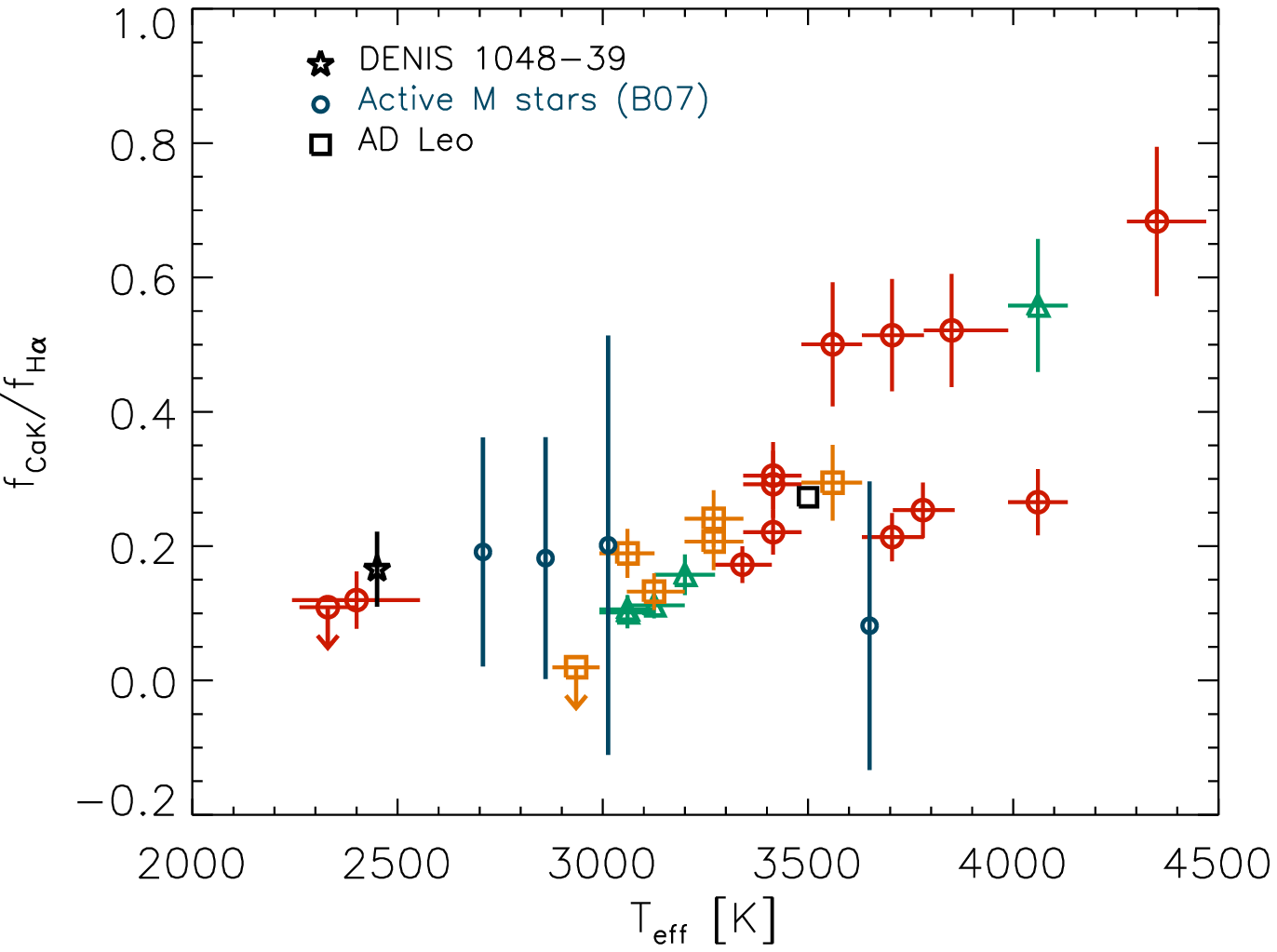}
}
}
\caption{Chromospheric line flux ratios versus effective temperature: 
{\it Left panel} -- H$\alpha$/H$\beta$,
{\it right panel} -- \ion{Ca}{\sc II}\,K/H$\alpha$. 
Same plotting symbols as in Fig.~\ref{fig:teff_spt_rotfit}. 
For comparison we include results for 
active M dwarf templates from the SDSS \protect\cite[blue circles;][]{Bochanski07.1}, 
the M9 dwarf DENIS\,1048-3956 \protect\citep[star symbol;][]{Stelzer12.0}, 
the PMS M dwarf AU\,Mic \protect\citep[black circle;][]{Houdebine94.2}, 
the M3.5 dwarf AD\,Leo \protect\citep[black square;][]{Mauas94.0},
and the Sun \protect\citep[rightward pointing arrows;][]{TandbergHanssen67.0}.}
\label{fig:decr_teff}
\end{center}
\end{figure*}

\subsection{Line decrements}\label{subsect:results_decr}

Our sample allows us to quantify the relative amount of emission in different parts
of the chromosphere for PMS stars throughout the whole M spectral class 
by means of the observed line flux ratios. Of particular interest are 
the Balmer decrements (e.g. $f_{\rm H\alpha}$/$f_{\rm H\beta}$) and the ratio
of the \ion{Ca}{\sc ii}\,K and H$\alpha$ flux. 
We preferred to examine the decrements as a function
of $T_{\rm eff}$ instead of SpT because the temperature is a physical parameter, 
while the SpT is merely an observational proxy of it.

In Fig.~\ref{fig:decr_teff} we compare flux ratios of the aforementioned 
emission lines with results for other stars and for the Sun that we 
retrieved from the literature. In the left panel the   
observed $f_{\rm H\alpha}/f_{\rm H\beta}$ ratios are shown together with  
the decrements for the active field M star templates defined by \cite{Bochanski07.1}, 
for the M9 field dwarf \object{DENIS-P\,J104814.7-395606} (henceforth DENIS\,1048-3956) 
derived by \cite{Stelzer12.0} on the basis of an X-Shooter spectrum, 
for the active dM3.5e star AD\,Leo \citep[see][]{Mauas94.0}, and 
for the PMS early-M dwarf AU\,Mic from \cite{Houdebine94.2}. This last one is particularly 
relevant because it is a member of the $\beta$\,Pic group with an age of $\sim 12$\,Myr
\citep{Zuckerman01.2, Kalas04.0}, similar to the age of the TWA. 
The typical H$\alpha$/H$\beta$ flux ratios of solar prominences and of
the solar chromosphere from \cite{TandbergHanssen67.0} are also indicated. 
For both the Class\,III objects and the field dwarfs, 
an increase of the H$\alpha$/H$\beta$ decrement towards lower $T_{\rm eff}$, that is  
late-M types, is evident.
However, due to the decrease of line fluxes the uncertainties at the end of the
M sequence are substantial. 

In the right panel of Fig.~\ref{fig:decr_teff} we show the 
$f_{\rm \ion{Ca}{\sc ii}\,K}/f_{\rm H\alpha}$ ratio of the Class\,III sample together with
the values observed for \object{DENIS1048-3956} and AD\,Leo,
%an active dM3.5e star studied by \cite{Mauas94.0}, 
and the results for the SDSS sample 
from \cite{Bochanski07.1}. For this latter one 
we propagated the errors given by 
\cite{Bochanski07.1}, who provided both lines with respect to H$\beta$. The resulting 
uncertainties are so large that the line ratios are consistent with zero. The measurements 
for our Class\,III sample are more meaningful. 
The observed \ion{Ca}{\sc ii}\,K/H$\alpha$ ratio of AD\,Leo
is consistent with that of the Class\,III stars of the same $T_{\rm eff}$.
We see a clear trend in our data for decreasing 
\ion{Ca}{\sc ii}\,K versus H$\alpha$ flux ratio towards lower $T_{\rm eff}$. 
The difference between the line flux ratio of early-M and late-M stars is much larger for
\ion{Ca}{\sc ii}\,K/H$\alpha$ (about a factor six from $T_{\rm eff} \sim 3000$\,K to 
$T_{\rm eff} \sim 4000$\,K) than for H$\alpha$/H$\beta$ (about a factor two in the same 
temperature range). 

A few features in the behavior of these flux ratios require more detailed investigation. 
First, the two stars, \object{TWA-6} and \object{TWA-14}, stand out from 
the trend in Fig.~\ref{fig:decr_teff}. They violate the smooth behavior of the flux ratios 
with spectral type in both panels. 
For these two objects the H$\alpha$ line is stronger than expected 
from their H$\beta$ and \ion{Ca}{\sc ii}\,K emission 
considering the trend of the other stars of similar effective temperature pointing at
differences in the structure of their outer atmospheres. 
Secondly, it is not clear if the trends described above are continuous throughout  
the lowest effective temperatures ($\lsim 3000$\,K). Especially the 
\ion{Ca}{\sc ii}\,K/H$\alpha$ ratio is  
sampled by only two PMS BDs, \object{TWA-26} and \object{TWA-29}, and by one field
ultracool dwarf, \object{DENIS1048-3956} given that the error bars on the SDSS data
are prohibitively large.  

\begin{figure}
\begin{center}
%\parbox{18cm}{
%\parbox{9cm}{
%\includegraphics[width=9cm]{./ps/HnHb_n.ps}
%}
\parbox{9cm}{
%HnHg_n.ps
\includegraphics[width=9cm]{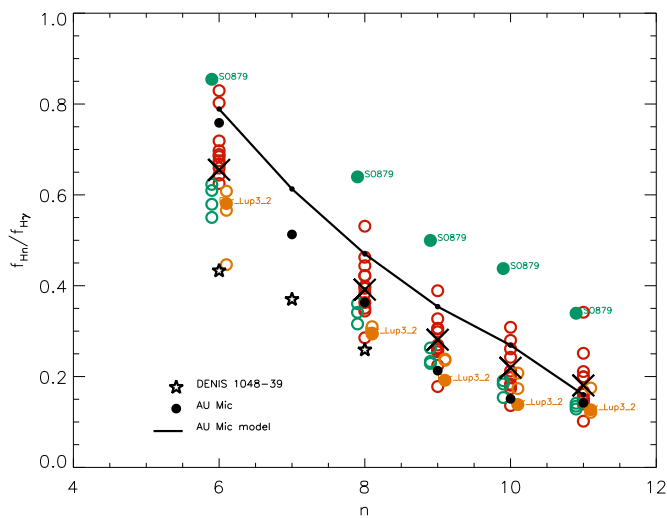}
}
%}
\caption{Balmer decrements with respect to H$\gamma$ for all stars from the Class\,III 
sample where the two lines comprising the decrement are detected. 
Same plotting symbols as in Fig.~\ref{fig:teff_spt_rotfit}. 
Predicted values from a 
chromospheric model devised for AU\,Mic are shown as a solid line, and the observed values
of AU\,Mic are represented by filled black circles.}
\label{fig:balmerdecr}
\end{center}
\end{figure}

In Fig.~\ref{fig:balmerdecr} we present the Balmer decrements up to $n=11$.
For clarity, members of the three different SFRs have been given a small horizontal offset
with respect to each other. 
Only stars with detections in the two lines that define the decrement are plotted,
and their mean value is shown as a large cross for each energy level $n$. 
We chose here to refer the decrements to H$\gamma$ because this allows us to 
compare our observations with results from the literature. The small circles connected with
a solid line represent the decrements derived by \cite{Houdebine94.2} from NLTE radiative
transfer modeling of the chromosphere of AU\,Mic. 
The observed values for this star are shown as larger filled black circles. 
%AU\,Mic is a member of the $\beta$\,Pic group with an age of $\sim 12$\,Myr 
%\citep{Zuckerman01.2, Kalas04.0}. 
An effective temperature of $3500$\,K and $\log{g}=4.75$ were assumed by \cite{Houdebine94.2} 
for the photosphere underlying the chromospheric modeling of AU\,Mic. 
Therefore, this model can be
expected to approximate the chromospheres of our Class\,III sample, and in particular the
TWA members among them. 
The model of \cite{Houdebine94.2} somewhat overpredicts their Balmer decrements, but the
general agreement is good.  
A tendency of larger decrements for hotter stars can be seen in Fig.~\ref{fig:balmerdecr}. 
We have highlighted 
two representatives of the hot and cool end of our sample, SO\,879 (K7) and Par-Lup3-2 
(M5). A detailed study of the temperature and gravity dependence of the Balmer decrements
is deferred to a later point when a grid of NLTE models is available. 

\section{Summary and conclusions}\label{sect:summary}

We have determined the photospheric parameters $T_{\rm eff}$ and $\log{g}$, 
rotation velocities, and radial velocities for $24$ Class\,III objects from the TWA,
$\sigma$\,Ori, and Lupus\,III star-forming regions using the ROTFIT routine to compare 
X-Shooter spectra with synthetic BT-Settl spectra in carefully selected wavelength
ranges. The values we found agree excellently with previous measurements
given in the literature where available. For four targets in Lupus, all five in 
$\sigma$\,Ori, and the two coolest of the TWA object(s) 
we presented the gravity, rotation, and kinematic parameters for the first time.
For the remaining two targets in Lupus, Sz\,121 and Sz\,122, we found a very fast rotation 
consistent with previous measurements. These two stars are suspected to be spectroscopic 
binaries. 

There are discrepancies between the gravities derived by us from the X-Shooter spectra
and those predicted by the evolutionary models of \cite{Baraffe98.1} and \cite{Chabrier00.2}. 
Our findings confirm
previous results. In particular, the coolest M-type objects, in our case the two BDs
\object{TWA\,26} and \object{TWA\,29}, have a lower gravity than the mid- to early-M stars,
a trend that is not present in the models. \cite{Mohanty04.2} have discussed 
problems with the initial conditions, effects of accretion, and the treatment of convection 
in the evolutionary models as possible causes for this mismatch. 
The region around
$T_{\rm eff} \sim 2500$\,K is characterized by the onset of dust formation, and it seems
that problems with the dust treatment within the models 
are responsible for the discrepant gravities across the M spectral sequence in TWA. 
No evolutionary models are available so far for the latest synthetic atmosphere grid, BT-Settl. 

The equivalent widths of the \ion{Li}{\sc i}$\lambda 6708$\,\AA\, line are similar
to previous estimates by \cite{Mentuch08.0} for the TWA members. No previous reports
of lithium exist for the $\sigma$\,Ori and Lupus objects from this sample. 
The comparison with the curves of growth from the literature 
\citep{Palla07.0, ZapateroOsorio02.2} yielded a rough estimate for the lithium abundances. 
We inferred approximate A(Li) $\approx 2.5 - 3.5$.  
Our lithium measurements are thus qualitatively consistent with the young age of the 
stars. 

We measured the activity indices, defined as ratio between the luminosity
in the activity diagnostic and the bolometric luminosity, %$R_{\rm H\alpha}$ and $R_{\rm X}$ 
and compared them with those of M dwarfs in the field. 
For the Class\,III sources of early-M spectral type the mean activity indices of 
H$\alpha$ ($\log{(L_{\rm H\alpha}/L_{\rm bol})} = -3.72 \pm 0.21$) 
and X-rays ($\log{(L_{\rm X}/L_{\rm bol})} = -2.85 \pm 0.36$) 
differ by almost one order of magnitude. Both the absolute values for the mean X-ray and H$\alpha$
activity indices and the differences between them are similar to those of the active ones among the 
early-M type field stars presented in the literature \citep[see e.g.][]{Stelzer13.0}. 
`Active' field dwarfs are generally those that rotate faster, which 
is likely due to relatively young age. Individual ages are generally not available
for field dwarfs, but in any case the younger ones among them are on the order of a
few $100$\,Myr old. Therefore, our finding indicates that
in M stars the decline of magnetic activity sets in only after the stars have reached
the main-sequence. Similar results have been obtained by \cite{Preibisch05.3} 
in a comparative study of X-ray luminosity functions from the Orion Nebula Cluster
($\sim 1$\,Myr), the Pleiades ($\sim 100$\,Myr), and the Hyades ($\sim 600$\,Myr) cluster, 
where the authors noted that the decay of activity with age
is slower for M stars than for G- and K-type stars. 
Both the H$\alpha$ and the X-ray activity index drop sharply for late-M spectral types.
Our sample is too small to determine the subtype at which the decline sets in. 
A similar decrease of H$\alpha$ and X-ray emission is observed in samples of field 
stars \citep[e.g.][]{West04.1}. It has been
ascribed to the increasingly neutral atmospheres of the coolest M dwarfs that impede
efficient coupling between matter and magnetic field \citep{Mohanty02.1}. 

No correlation with rotational velocity is seen in the Class\,III sample, indicating
that the activity of these stars is in the saturated regime, consistent with previous
studies of X-ray emission in PMS stars \citep{Preibisch05.1}. The situation is less clear 
for the two BDs, 
\object{TWA\,26} and \object{TWA\,29}. Their rotational velocities and stellar radii 
indicate an upper limit to their rotation period of $\sim 8$\,h. This value is within
the range of $4.1$ to $88$\,h for photometrically observed periods of young BDs 
in the $5$\,Myr old $\epsilon$\,Ori cluster \citep{Scholz05.1}. In spite of their evidently
fast rotation \object{TWA\,26} and \object{TWA\,29} have activity levels (in terms of
the activity indices) that are at least one order of magnitude lower than those of early-M
Class\,III objects. 
While the decline of H$\alpha$ activity index at the cool end of the stellar sequence
is well-established for evolved field dwarfs \citep[see e.g.][]{West04.1, Bochanski07.1}, 
the PMS H$\alpha$ index has to the best of our knowledge not been 
systematically explored down to the substellar limit. 
%The chromospheric saturation level
%of $\log{(L_{\rm H\alpha}/L_{\rm bol})} = -3.3$ defined 
%by \cite{Barrado03.2} yields an approximate upper limit for the chromospheric H$\alpha$
%emission. However, it has been determined from $\sim 100$\,Myr-old stars of early-M spectral 
%type and may not represent the typical chromospheric level of younger and cooler objects. 
On the X-ray side, even the most sensitive available star-forming region 
surveys such as the Chandra Orion Ultradeep field \citep[cf.][]{Getman05.1} are barely deep
enough for the detection of BD coronae. An X-ray study of the $3$\,Myr-old 
cluster IC\,348 has yielded a slope steeper than 
unity for the relation between $L_{\rm x}$ and $L_{\rm bol}$, corresponding to a decrease 
of the activity index with decreasing luminosity, that is for cooler objects \citep{Stelzer12.1}. 
However, the significance of this result is drastically reduced when only Class\,III sources 
are considered, suggesting that accretion and not the chromosphere is at the origin of
this trend. To summarize, the low activity indices seen in the two Class\,III BDs 
of the TWA need to be bolstered with systematic studies of larger samples of young 
substellar chromospheres. If confirmed, 
this may indicate that in the substellar regime the dynamo efficiency decouples   
from the rotation rate, or that some other process such as disruption of the corona by
centrifugal forces \citep{Jardine04.0} suppresses the emission. 

We estimated the relative contribution of the chromosphere and the corona to the
radiative output produced by magnetic activity by comparing the observed fluxes.
For the chromosphere the fluxes of all emission lines in the X-Shooter spectrum were 
summed, while for the corona we used the soft X-ray flux. 
In our sample the X-ray fluxes tend to be higher than the optical radiation measured 
by the emission lines. Observations extracted from the literature 
suggest that the UV emission is negligible for the chromospheric radiation budget of these 
stars, but this has to be corroborated once larger samples of Class\,III sources 
have sensitive constraints on their UV fluxes. 
The spread for the flux
ratio $F_{\rm X}/F_{\rm opt}$ is at least one order of magnitude in our Class\,III
sample and a trend for an age dependence of the coronal-to-chromospheric flux ratio
is suggested, but requires a larger sample size to be firmly established. 
Note that the X-ray and optical observations are not simultaneous. The influence
of variability, therefore, remains unknown. 

Flux-flux relations between individual chromospheric emission lines and those between
chromospheric diagnostics and X-ray emission were presented. The results for the 
Class\,III stars confirm the existence of the upper `active' branch identified 
by MA\,11 on the
basis of a few objects in a large sample of field FGKM stars that deviated from the 
flux-flux relations of the bulk of objects. As far as we can say considering our sample
of $24$ Class\,III sources, all young stars populate this `active' branch. 
The `active' stars are characterized by a higher ratio between H$\alpha$ and
\ion{Ca}{\sc ii} emission (both \ion{Ca}{\sc ii}\,K and the IRT) 
than that of the less active ones. Similarly, their X-ray emission
is enhanced with respect to \ion{Ca}{\sc ii} emission. 
Neither MA11 nor our study found differences in the $F_{\rm H\alpha}$ vs. $F_{\rm X}$ 
relation for stars from the two branches.   
Since all these diagnostics are formed in different atmospheric layers,  
our finding indicates a change in the response of the atmosphere to the drivers of magnetic 
activity as the activity decreases. This is presumably a consequence of rotational
evolution and the ensuing decay of dynamo efficiency. Flux-flux relations for samples
known to rotate slowly are needed for a comparison with the very active Class\,III stars. 
In particular, it seems that there is a clear distinction
between low-activity stars and those with strong H$\alpha$ and X-ray emission. 
Given the relation between activity and age, the relative amount of emission in the
various diagnostics may, therefore, in the future be applied as a qualitative youth indicator. 

Within the spectral class M, 
the flux ratio between \ion{Ca}{\sc ii}\,K and H$\alpha$ emission shows a significant decrease
towards lower $T_{\rm eff}$, while the H$\alpha$/H$\beta$ decrements are slightly increasing but 
are less sensitive to the effective temperature. 
This may again be a signature of the different layers where these lines form. 
Two stars deviating from these trends, \object{TWA\,6} and \object{TWA\,14}, are the
fastest rotators in our sample. At the same H$\beta$ and \ion{Ca}{\sc ii}\,K levels of
other stars with similar effective temperature they present higher H$\alpha$ flux.
We conclude that H$\alpha$ is more sensitive to rotation than other emission lines. This
is probably the reason for the existence of the two branches in the flux-flux relations
discussed above. 
No chromospheric model grids are 
available for the parameter space covered by our Class\,III sample to be compared
with our observations. A dedicated H$\alpha$
and \ion{Ca}{\sc ii} study was presented by \cite{Houdebine97.1} for much less active dM1 
stars but no theoretical line fluxes were provided. 

The series of Balmer decrements from H$\delta$ to H11 referred to H$\gamma$ for our
Class\,III sample 
are roughly 
reproduced with the NLTE radiative transfer model devised for the young M dwarf AU\,Mic by 
\cite{Houdebine94.2}. The trend is towards smaller decrements for later-M spectral types,
analogously to the result for the inverse of H$\alpha$/H$\beta$. 
Our findings represent a guideline for future chromospheric modeling efforts. Such models 
should be able to predict the behavior of line flux ratios throughout the whole M
spectral class. Moreover, they should explain the emissions of stars with different activity levels.

\begin{acknowledgements}
We thank an anonymous referee. 
JMA and BS recognize the active contribution of G.Attusino to some of the teleconferences.
We thank the ESO staff for their support during the observations. 
We also appreciate the support of P. Goldoni, A. Modigliani, and G. Cupani 
in the use of the X-Shooter pipeline. 
\end{acknowledgements}

\bibliographystyle{aa} %aa.bst
\bibliography{chromo}

\end{document}